\documentclass[useAMS,usenatbib,onecolumn]{mn2e}
\usepackage{amsmath}
\usepackage{amssymb}
\usepackage{graphicx}
\usepackage{euscript}
\usepackage{bm,tabularx}
\usepackage{float,color}
\graphicspath{{./figs/}}

\newcommand{\beq}{\begin{equation}}
\newcommand{\eeq}{\end{equation}}

\newcommand{\bfa}{\mbox{\boldmath $a$}}

\newcommand{\bfc}{\mbox{\boldmath $c$}}

\newcommand{\bfk}{\mbox{\boldmath $k$}}

\newcommand{\bfq}{\mbox{\boldmath $q$}}
\newcommand{\bfu}{\mbox{\boldmath $u$}}
\newcommand{\bfv}{\mbox{\boldmath $v$}}
\newcommand{\bfx}{\mbox{\boldmath $x$}}
\newcommand{\bfy}{\mbox{\boldmath $y$}}

\newcommand{\bfA}{\mbox{\boldmath $A$}}
\newcommand{\bfB}{\mbox{\boldmath $B$}}
\newcommand{\bfC}{\mbox{\boldmath $C$}}

\newcommand{\bfH}{\mbox{\boldmath $H$}}
\newcommand{\bfK}{\mbox{\boldmath $K$}}

\newcommand{\bfU}{\mbox{\boldmath $U$}}

\newcommand{\bfX}{\mbox{\boldmath $X$}}
\newcommand{\bfY}{\mbox{\boldmath $Y$}}
\newcommand{\bfQ}{\mbox{\boldmath $Q$}}

\newcommand{\ex}{\mbox{{\boldmath $e$}}_{1}}
\newcommand{\ey}{\mbox{{\boldmath $e$}}_{2}}
\newcommand{\ez}{\mbox{{\boldmath $e$}}_{3}}

\newcommand{\bnabla}{\mbox{\boldmath $\nabla$}}
\newcommand{\cross}{\mbox{\boldmath $\times$}}
\newcommand{\cendot}{\mbox{\boldmath $\cdot\,$}}

\newcommand{\bsy}{\boldsymbol}
\newcommand{\ud}{\mathrm{d}}
\newcommand{\Eq}[1]{Eq.~(\ref{#1})}

\newcommand{\Eqss}[2]{Eqs.~(\ref{#1})--(\ref{#2})}

\newcommand{\Fig}[1]{Figure~\ref{#1}}
\newcommand{\Figs}[2]{Figures~\ref{#1} and~\ref{#2}}

\definecolor{dred}{rgb}{.9, .15,.1}
\definecolor{brown}{rgb}{0.5,0.2,0.2}
\definecolor{dgreen}{rgb}{0.15,0.6,0.15}
\definecolor{dblue}{rgb}{0.0,0.0,0.6}

\begin{document}

\title[Mean field dynamo action in shear flows]
{Mean field dynamo action in shear flows. I: fixed kinetic helicity}

\author[Jingade \& Singh]
{Naveen Jingade$^{1,2}$\thanks{E-mail: naveen@rri.res.in}
and Nishant K. Singh$^{3,4}$\thanks{E-mail: nishant@iucaa.in}\\
 $^{1}$Indian Institute of Science, Bangalore 560 012, India\\
 $^{2}$Raman Research Institute, Sadashivanagar, Bangalore 560 080, India\\
 $^{3}$Max Planck Institute for Solar System Research,
Justus-von-Liebig-Weg 3, D-37077 G\"ottingen, Germany\\
 $^{4}$Inter-University Centre for Astronomy and Astrophysics,
 Post Bag 4, Ganeshkhind, Pune 411 007, India
}

\pagerange{\pageref{firstpage}--\pageref{lastpage}} \pubyear{}

\maketitle

\label{firstpage}

\begin{abstract}
We study mean-field dynamo action in a background linear shear flow by
employing pulsed renewing flows with fixed kinetic helicity and
nonzero correlation time ($\tau$). We use plane
shearing waves in terms of time-dependent exact solutions to
the Navier-Stokes equation as derived by \cite{SS17}.
This allows us to self-consistently include the anisotropic
effects of shear on the stochastic flow. We determine the average
response tensor governing the evolution of mean magnetic field,
and study the properties of its eigenvalues which yield the growth
rate ($\gamma$) and the cycle period ($P_{\rm cyc}$) of the mean
magnetic field. Non-axisymmetric mode of the mean-field decay as $t\to \infty$ and hence are deemed unimportant for mean-field dynamo. Both, $\gamma$ and the wavenumber corresponding
to the fastest growing axisymmetric mode vary non-monotonically
with shear rate $S$ when $\tau$ is comparable to the eddy turnover
time $T$, in which case, we also find quenching of dynamo when shear
becomes too strong. When $\tau/T\sim{\cal O}(1)$, the cycle
period ($P_{\rm cyc}$) of growing dynamo wave scales with shear as
$P_{\rm cyc} \propto |S|^{-1}$ at small shear, and it becomes
nearly independent of shear as shear becomes too strong.
This asymptotic behaviour at weak and strong shear has
implications for magnetic
activity cycles of stars in recent observations.
Our study thus essentially generalizes the standard
$\alpha \Omega$ (or $\alpha^2\Omega$) dynamo as also the $\alpha$ effect is
affected by shear and the modelled random flow has a finite memory.
\end{abstract}


\section{Introduction}
Coherent large-scale magnetic fields and mean differential
rotation are two common features of most astrophysical objects,
such as, the Sun, stars, galaxies, etc
\citep{Par79,ZRS83,RSS98,Jon11,Han17}.
Magnetic fields in these systems are maintained by turbulent
dynamo action where the standard paradigm for large-scale
component involves amplification of weak seed fields due to
helical turbulence in shear flows \citep{Mof78,KR80,BS05}.
The $\alpha$ effect, which, in idealized settings,
is a measure of net kinetic helicity
and arises naturally in systems with rotation and stratification,
plays a crucial role in driving large scale dynamos in a
variety of systems \citep[e.g.][]{DS07,Cha10,BSS12}; see also
\cite{CHT06,HY08} for a more general description of the
$\alpha$ effect, where non-local and non-instantaneous
effect are considered in the expression of turbulent EMF by
employing integral relation between EMF and mean-field,
unlike simplified standard dynamo models where the 
EMF depends on the local and instantaneous value of the
mean magnetic field.

Direct numerical simulations in galactic or solar contexts
have shown that the large scale magnetic fields are
naturally produced as a result of a mean-field turbulent
dynamo in local as well as global setups, where the mirror
symmetry of turbulence
is broken either by having a helical driving or
by enabling convection in a rotating system
\citep{BBS01,GEZR08,KMB12,WKKB14,Kapy16,KGVS18}.
Mean shear is common to these studies and the dynamo
is thought to be of $\alpha \Omega$-type, or,
more generally, of $\alpha^2 \Omega$-type as the role
of $\alpha$-term in the generation of shear-wise component
may indeed be comparable to the $\Omega$-effect
\citep{VKWKR19};
see, e.g., \cite{BS05}, for different types of dynamos.
It is known that the $\alpha$-effect is a more complicated tensorial
object \citep{RKR03,BS05}, which might be much different
from the net kinetic helicity of the flow \citep{KR03,CS17}.
Turbulent transport coefficients, such as the $\alpha$-tensor,
are often numerically determined in a variety of contexts
\citep{SBS08,MKTB09}. 

Somewhat less common approach to study the large-scale
dynamo is to directly solve for the evolution of mean
magnetic field by determining the response function for
a given smooth random flow \citep{GB92,KSS12}.
Such a model is shown to faithfully represent a fast
or small-scale dynamo \citep{ZRMS84,BC88,GB92}.
Essentially all astrophysical bodies are expected to host a
small-scale dynamo, which appears to be always
excited when the magnetic Reynolds number exceeds a
critical value \citep{ZRS83}. There are concerns that its
presence makes the description of large-scale dynamo in
terms of standard mean field magnetohydrodynamics (MHD),
or, say, $\alpha$-prescription,
less straightforward \citep[e.g.][]{SBK02,CHT06}.

\cite{GB92}, hereafter GB92, chose random helical flows in their model
where they also included the memory effects, and
showed analytically that the magnetic field
develops intermittency in time\footnote{The term `memory effect'
is used in this work to indicate that the random flow has non-zero
correlation times. As clarified in \cite{SS14}, this is equivalent to
more usual notion of memory effects in dynamo theory when the turbulent EMF is
affected by the time dependence of the mean magnetic field, whereas
in the white-noise case, the generalized EMF
with a history term through a time-integral reduces to a simple expression
leading to an instantaneous relation with the mean magnetic field.}.
They also found growing
solutions for the first moment or the mean magnetic field,
and, in the limit of small correlation times, they
recovered the predictions of standard $\alpha^2$ dynamo
for the growth rate. \cite{KSS12}, hereafter KSS12,
extended the work of GB92
to also include the effect of shear, and showed quite generally,
that the mean field dynamo action is not possible so long as
the flows are strictly non-helical. By considering a particular
model of random helical flow, they derived
a generalized response tensor, which yields,
for fixed kinetic helicity and small correlation times,
the same dispersion relation as that from a standard
$\alpha^2 \Omega$ dynamo. In paper~II \citep{JS20}
we consider flows with fluctuating kinetic helicity,
which has a renovation time greater than the velocity
renovation time, and show that the growth of mean-magnetic
field is possible even in the absence of negative
turbulent diffusion \citep{Kra76}.  

Following GB92; KSS12, we adopt here a renovating flow
based model which allows us to describe the evolution of mean
magnetic field, without explicitly deriving any mean electromotive
force (EMF) which is essential in standard mean field MHD.
In this model, one assumes an exactly solvable flow field in
terms of a single plane wave which renovates itself after each
time interval $\tau$.
Thus, the time is split into equal intervals of length $\tau$
and the velocity fields in different intervals are assumed to be
statistically independent. The evolution of magnetic field
is then determined by the realization of the velocity field
only in a fixed interval.
Such a model neither involves any closure approximation, nor
is it limited to low fluid or magnetic Reynolds numbers.
KSS12 in their analysis considered forced overdamped shearing waves
to model the renovating flows.
Here we make use of plane shearing waves, which are time-dependent
exact solutions to the Navier-Stokes equations as
derived by \cite{SS17}. Shear induces anisotropy in the
renovating flow, which, in this work, is allowed to freely
decay for the renovation time $\tau$, and it is resetted
to the same amplitude at the beginning of each renovation
interval. Such a resetting essentially tries to capture
the effects of a random forcing after every $\tau$.

In Section~\ref{model} we present the model of renovating flow
in a background linear shear flow and then discuss the helical
shearing waves which are used in this work.
Linear shear makes the induction equation inhomogeneous in
the lab frame; see \Eq{InducSh1}. To deal with this, we make use of shearing
coordinate transformation
and then determine the Cauchy's solution for ideal induction
equation in shearing frame. We then derive an expression for
the average response tensor governing the evolution of mean magnetic field
in Fourier space after suitably averaging over
randomness of the flow. Behaviour of non-axisymmetric modes of mean magnetic
field is presented in Section~\ref{nonaxisy}.
In Section~\ref{axi} we present our findings on the axisymmetric mean field
dynamos, where we explore various properties of dynamo growth rate and its
cycle period in detail, and discuss the significance of new predictions from
our model in light of recent observations. We conclude in Section~\ref{con}.

\section{model description}
\label{model}

We, now investigate the evolution of the mean-magnetic field in the background shear flow along with the turbulence. Such systems are common in astrophysical scenario like Sun, Galaxies, accretion disk etc. Let ($\ex,\ey,\ez$) be the orthonormal unit vectors in the Cartesian coordinate system in the lab frame, where $\bfX=(X_1,X_2,X_3)$ is a position vector.  We choose mean shear to be acting in the $\ey$ direction (azimuthal direction),  varying along $X_1$ linearly, which is a local shearing--sheet approximation to the differential rotation of the disks \citep{GL65,BRRK08}. The model velocity field $\bfU$ can be written as, $\bfU(\bfX,t) = S\,X_1\ey + \bfu(\bfX,t)$, where $\bfu$ is the turbulent velocity field, and the shear rate $S$ is a constant parameter.

\subsection{Renewing flows in shearing background}  
Let us consider the inviscid Navier--Stokes equation in the background linear shear flow for the unit mass density,
\beq
\left(\frac{\partial}{\partial t}+SX_1\frac{\partial}{\partial X_2}\right)\bfu+Su_1\ey +(\bfu\cendot\bnabla)\bfu=-\bnabla p\,, \quad\mbox{with}\quad \bnabla\cendot\bfu = 0, 
\label{NSeqn}
\eeq
where we have also assumed the flow $\bfu$ to be incompressible.
We look for the single helical wave solution of the form,  
\beq
\bfu(\bfX,t) = \bfA(St,\bfq)\sin(\bfQ(t)\cendot\bfX + \Psi) + h\,\bfC(St, \bfq)\cos(\bfQ(t)\cendot\bfX + \Psi)
\label{vel}
\eeq
where $\bfQ(t)$ is a shearing wavevector having the form
$\bfQ = (q_1-S q_2(t-t_0),q_2,q_3)$, $\bfq = (q_1,q_2,q_3)$ is the
wavevector at initial time $t_0$, and $\Psi$ denotes the phase of the wave.
This particular form of wave vector arises because of the inhomogeneity of the \Eq{NSeqn} in the variable $X_1$. $\bfA(St,\bfq)$ and $\bfC(St,\bfq)$ are the amplitudes of the sheared helical wave and $h$ controls the relative helicity of the flow. The above velocity field is supplemented by 
\beq
\bfQ(t)\cendot\bfA(t) = 0; \quad \mbox{and} \quad \bfQ(t)\cendot\bfC(t) = 0; 
\label{IncomCon}
\eeq
because of the incompressibility condition of velocity field. This also leads to the constancy of phase of the wave i.e., $\bfQ(t)\cendot\bfX = \bfq\cendot\bfx_0$, where $\bfx_0$ is initial position of the fluid particle. Because of this, we can easily integrate (either numerically or analytically) the \Eq{vel} to obtain the Lagrangian trajectory of the fluid particle, later to be used in the Cauchy's solution given in \Eq{Cauchy}, which is an integral equation.
Such single scale flows are used in many studies exploring
the intermittency, small-scale and large-scale dynamos
\citep{GB92,BS15,KSS12}. 
One of the advantages of such a procedure is that
it bypasses the closure schemes which are somewhat limiting
the mean-field dynamo theories. 

\subsubsection{Shearing waves}
When we substitute \Eq{vel} in \Eq{NSeqn}, the non linear term $(\bfu\cendot\bnabla)\bfu$ vanishes, whereas the term $SX_1(\partial \bfu/\partial X_2)$ is nonzero describing
the interaction of large scale motion (background shear) with the turbulent velocity field $\bfu$. Therefore, we get time-dependent wave vectors and amplitude modulation by shear in the helical wave as shown in \Eq{vel}.
We adopt the following expression for the velocity amplitudes ($\bfA$, $\bfC$) which
were derived in \cite{SS17}:
\begin{eqnarray}
[A_1(\bfq, t),C_1(\bfq, t)] &=& \frac{q^2}{Q^2(t)}\,
[a_1,c_1]\,,
\label{solna1es}\\[2em]
[A_2(\bfq, t),C_2(\bfq, t)] &=& [a_2,c_2] \;+\;\frac{q^2}{q_{\perp}^2} \left(\frac{q_3^2}{q_2\, q_{\perp}}{\cal M}(\bfq,t)\;-\; q_2\,{\cal N}(\bfq,t)\right)\,[a_1,c_1]\,,
\label{solna2es}\\[2em]
[A_3(\bfq, t),C_3(\bfq, t)] &=& [a_3,c_3] \;-\; \frac{q^2\,q_3}{q_{\perp}^2}\left(\frac{{\cal M}(\bfq,t)}
{q_{\perp}}\;+\; {\cal N}(\bfq,t)\right)\,[a_1,c_1]\,,
\label{solna3es}
\end{eqnarray}
where
\begin{eqnarray}
{\cal M}(\bfq,t) & = & \arctan\left(\frac{Q_1(t)}{q_{\perp}}\right) \,-\,\arctan\left(\frac{q_1}{q_{\perp}}\right)\quad \mbox{and} \quad {\cal N}(\bfq,t)  = \frac{Q_1(t)}{Q^2(t)} \;-\; \frac{q_1}{q^2} \nonumber
\end{eqnarray}
where $(\bfa,\bfc)$ are amplitudes of the velocity field at initial time $t_0$. The vectors ($\bfq$, $\bfa$, $\bfc$) form an orthogonal triad. The time-dependent wave vector
is given by 
\beq
Q^2(t) = Q_1^2(t) + q_{\perp}^2; \quad Q_1(t) = q_1-S(t-t_0)q_2, \quad q_{\perp}^2 = q_2^2 + q_3^2.
\eeq
These solutions represent the local disturbance of the velocity field in shear flows. The amplitudes $(A_1,C_1)$ decrease with time, whereas $(A_2,C_2)$ and $(A_3,C_3)$ increase with time and then saturate. These helical--sheared waves rotate towards $X_1$-direction (or negative $X_1$-direction, depending on the initial value of $\bfq$) in the $X_1-X_2$ plane as they propagate, due to the dependency of wave vector component $Q_1$ on shear \citep[see,][for details]{SS17}.

The helicity $H$ of the turbulent velocity field is defined as (following KSS12),
\beq
H = h\,\bfC(t)\cendot\left(\bfQ(t)\cross\bfA(t)\right),
\label{hel}
\eeq
The parameter $h$ takes values in the range $[-1,1]$ and it controls the strength of the helicity; $h=\pm 1$ corresponds to maximally helical flow. From the amplitudes given in \Eqss{solna1es}{solna3es}, it can be shown after straightforward and tedious algebra that $H=h\,\bfc\cendot(\bfq\cross\bfa)$. Even though the amplitudes $\bfA(St,\bfq)$ and $\bfC(St, \bfq)$ are function of shear $S$, helicity $H$ of the fluid particle in this flow is independent of shear $S$, which is rather intriguing.

Let us construct the random flows using shearing waves, that we just introduced. In the renovating flow model, time is split into the equal intervals of length $\tau$. The time $\tau$ is defined as the renovation time of the random process $\bfu(\bfX,t)$. The velocity field over these intervals are assumed to be distributed randomly and independently. The statistical distribution of random flow is assumed to be invariant to the shift of shearing coordinate $\bfx$, a natural symmetry of shear flows \citep{SS11}. These distributions are also assumed to be constant over the intervals [$(n-1)\tau$,$n\tau$]; $n=1,2,3 \ldots$. Such ensembles simplify the dynamo problem considerably \citep{Kra76b,KR80}. These velocity fields are stationary over the discrete times $\tau$, $2\tau$, $3\tau$ $\ldots$. Hence, these can be approximated as a stationary random process over a long time ($\gg\tau$) with exponentially decaying time correlations \citep{Mol84}.
It is known that such velocity field together with ensemble can give rise to fast dynamo action \citep{FO88,Mol85}. 

We employ the same ensemble as in GB92, KSS12: where $\Psi$ is randomnly distributed from $0$ to $2\pi$, this preserves the homogeneity in the shearing co-ordinates\footnote{The velocity correlator are independent to the shift of the shearing coordinate $\bfx$:  $\left\langle \bfv_i(\bfX,t)\bfv_j(\bfY,t')\right\rangle=f_{ij}(\bfQ(t)\cendot\bfX-\bfQ(t')\cendot\bfY)=f_{ij}[\bfq\cendot(\bfx-\bfy)]$, where $f_{ij}$'s are some function.}, whereas in the absence of shear, it would give usual homogeneity condition; the direction of wave vector $\bfq$ is distributed randomly over the sphere of radius $q$, and this is assumed to take random direction in the successive intervals; $\bfa$ and $\bfc$ are distributed over the circle of radius $a$, perpendicular to the wave vector $\bfq$. At the beginning of every time interval, the wave vector $\bfQ$, and the amplitudes ($\bfA$, $\bfC$) are reset to it's initial values. 

\subsection{Evolution of mean-magnetic fields in renovating flows}
\label{form}
The evolution of the magnetic field in the background shear flow with the velocity field $\bfU(\bfX,t) = S X_1\ey + \bfu(\bfX,t)$ is given by
\beq
\left(\frac{\partial}{\partial t} + SX_1\frac{\partial }{\partial X_2}\right)\bfB +(\bfu\cendot\bnabla)\bfB-S\,B_1 \ey = (\bfB\cendot\bnabla)\bfu + \eta\nabla^2 \bfB
\label{InducSh1}
\eeq

As our interest is in the mean-magnetic field whose scale is much larger than
the energy injection scale of turbulence, we safely ignore the diffusion term in
\Eq{InducSh1}, in this work. We note, however, that this would eliminate the
threshold for the small-scale dynamo (SSD) in a kinematic study like this, and would lead
to the growth of smaller scale magnetic structures; see, e.g., \cite{Mol85,DUO93}
to also note that the growth rate of SSD is independent of microscopic resistivity
as $\eta\to 0$. Thus, while magnetic fields are expected to
be produced at small scales in each realization of the ensemble,
these small-scale structures would average to zero in the ensemble average,
by definition, which we adopt here to study only the mean magnetic field.
Henceforth, we focus only on the
evolution of the first moment of the magnetic field; the second moment which will be
suitable for studying the SSD will be explored elsewhere.

Equation~(\ref{InducSh1}) is inhomogeneous in the co-ordinate $X_1$, therefore, it can be best solved in shearing co-ordinates \citep{SS10}. The shearing transformation is obtained by solving
\beq
\frac{\ud \bfX}{\ud t} = SX_1\ey\, ,
\eeq
which gives, 
\beq
X_1 = x_1; \quad X_2 = x_2 + S(t-t_0)x_1; \quad X_3 = x_3\, , 
\label{shear-Trans}
\eeq
where $\bfx$ is the Lagrangian coordinate of fluid element carried along by the background shear flow, and $t_0$ is the initial time. We can write the above transformation in compact form, if we introduce $\gamma_{ij} = \delta_{ij} + S\,(t-t_0)\,\delta_{i2}\delta_{j1}$. Then, we can write \Eq{shear-Trans} as
\beq
X_i = \gamma_{ij}(t-t_0)\,x_j
\label{shear-gamma}
\eeq
Let us write \Eq{InducSh1} in this new Lagrangian coordinates $\bfx$ and time $s = t-t_0$. Also, let us introduce new vector functions for the magnetic field, $\bfH(\bfx,s) = \bfB(\bfX,t)$ and for the velocity field, $\bfv(\bfx,s) = \bfu(\bfX,t)$. Then \Eq{InducSh1} becomes, 
\beq
\frac{\partial \bfH}{\partial s}  +(\bfv\cendot\bnabla)\bfH-S\,H_1 \ey = (\bfH\cendot\bnabla)\bfv, \quad\mbox{with} \quad \bnabla\cendot\bfv = 0, \quad  \bnabla\cendot\bfB = 0\, ,
\label{InducSh2}
\eeq
\beq
\mbox{where} \quad \bnabla = \frac{\partial}{\partial \bfx}- S\,s\, \ex\frac{\partial}{\partial x_2} \quad \text{is a time dependent operator.} \nonumber
\eeq
Equation~(\ref{InducSh2}) differs by the original induction equation by the term $-S\,H_1\ey$. We can eliminate this term by transforming to a new magnetic field variable $h_j$
which is defined from $H_i = \gamma_{ij}(s)h_j$ (similar to transformation given in \Eq{shear-gamma}). Then we can write \Eq{InducSh2} in component form as:
\beq
\left(\frac{\partial}{\partial s} + [\gamma_{jk}(-s)v_k]\frac{\partial}{\partial x_j}\right)h_i = h_k\frac{\partial}{\partial x_k}[\gamma_{ij}(-s)v_j]
\label{InducSh3}
\eeq
Here we have used the property $\gamma_{ij}(s)\gamma_{jk}(-s)=\delta_{ik}$.
Equation~(\ref{InducSh3}) is similar to the induction equation except for the velocity field, where we have obtained in it's place, the transformed velocity field $v_j - Ss\,\delta_{j2}v_1$. We can then write the Cauchy's solution to \Eq{InducSh3} \citep{Lu51} as,  
\beq
h_i(\bfx,s) ={\cal J}_{ij}(\bfx,s)h_j(\bfx_0,0)
\eeq
where $\bfx = \bfx_0 + \int_0^s (\bfv-S s'\ey v_1)\ud s'$ and 
\beq
{\cal J}_{ij} =  \frac{\partial x_i}{\partial x_{0j}} = \delta_{ij}+\displaystyle \int_0^s \left(\frac{\partial v_i}{\partial x_{0j}}-S s'\delta_{i2}\frac{\partial v_1}{\partial x_{0j}}\right)\ud s'\,.
\label{Jacobian}
\eeq
Here $\bfx_0$ is the initial position of the particle at time $t=t_0$.
The magnetic field $H_i$ in the shearing frame:
\beq
H_i(\bfx, t) = \gamma_{ij}(t-t_0){\cal J}_{jk}(\bfx_0,t-t_0)H_{k}(\bfx_0,t_0)
\label{Cauchy}
\eeq
Since velocity fields are assumed to be uncorrelated on the successive intervals of the renovating flow model,
we can write the evolution of magnetic field in the general interval $[(n-1)\tau,n\tau]$ as, 
\beq
H_i(\bfx, n\tau) = \gamma_{ij}(\tau){\cal J}_{jk}(\bfx_0,\tau)H_{k}(\bfx_0,(n-1)\tau)
\label{Finalcauchy}
\eeq
We would now wish to calculate the response tensor for the average magnetic field. Since, we assume, the sheared renewing flow is homogeneous in the shearing coordinate $\bfx$, which is a natural symmetry in shear flows \citep[see,][]{SS11}, we can define the Fourier transform for the average of magnetic field $\left\langle H_i(\bfx, n\tau)\right\rangle$ in terms of the shearing waves as
\beq
\widetilde{H}_i(\bfk,t) = \int \left\langle H_i(\bfx, t)\right\rangle \exp(-i\bfk\cendot\bfx)\,\ud^3 x
\label{Fou-Trans2}
\eeq
where $\bfk= \bfK(t_0)$ is the initial wavevector at time $t_0$, which for each interval we take to be the time $(n-1)\tau$. Here the phase of the Fourier mode is conserved in time i.e., $\bfk\cendot\bfx=\bfK(t)\cendot\bfX$, where we take $t$ to be the time $n\tau$. Therefore, the relation between $\bfK(t_0)$ and $\bfK(t)$ is given by $k_i = \gamma_{ji}(t-t_0)K_j$ (see \cite{SS10,SS11} for details). We can see that the wavevectors depends on the interval of choice, which is the important destinction as compared to the case when the shear is absent, where wavevectors are time-independent. Substituting the \Eq{Finalcauchy} in \Eq{Fou-Trans2}, we get

\beq
\widetilde{H}_i(\bfk,n\tau) = \int \left\langle \gamma_{ij}(\tau){\cal J}_{jk}(\bfx_0,\tau)\overline{H_{k}(\bfx_0,(n-1)\tau)}\right\rangle_{\bfx} \exp(-i\bfk\cendot\bfx)\ud^3 x
\label{Fou-mag}
\eeq
Since we are in the kinematic regime, where the strength of initial magnetic field is assumed weak, there is no back reaction on the velocity field (no Lorentz force). In such a scenario, the velocity field statistics become independent of the statistics of the initial magnetic field. In the following, we carry averaging in two steps: first it is performed over the initial randomness of the magnetic field, which we denoted here by over-bar in \Eq{Fou-mag}; and second, it is carried over the statistical ensemble of the velocity field denoted by angle-brackets. This averaging is equivalent to averaging over the energy injecting scale of turbulence $q$ and it would introduce the effective turbulent diffusivity and smoothen the field over the scale $q$ \citep{Hoy87}. Hence, small scale magnetic structures would
vanish after the ensemble average and if there is any large scale structure,
it would reveal itself as a mean-field. The notation  $\left\langle\;\;\right\rangle_{\bfx}$ indicate that the averaging is carried over that trajectory, in each realization, which reaches a fixed point $\bfx$ at time $n\tau$.  By homogeneity of velocity field statistics in the shearing coordinate $\bfx$, the averaging becomes independent of spatial point $\bfx$. The initial magnetic field need not be homogeneous and its spatial dependency can be taken into account by the Fourier transform as defined by \Eq{Fou-Trans2}. Thus, using all the ingredients, we obtain\footnote{The mean-magnetic field in Fourier space at mode $\bfk$ has contribution only from the initial magnetic field at mode $\bfk$, which is an important simplification that is occurred, because of the homogeneity condition in the shearing coordinate $\bfx$.}

\beq 
\widetilde{H}_i(\bfk,n\tau) = {\cal G}_{ik}(\bfk,\tau)\widetilde{H}_{k}(\bfk,(n-1)\tau),
\label{MatrixEq}
\eeq

where

\beq
{\cal G}_{ik}(\bfk,\tau) = \left\langle \gamma_{ij}(\tau){\cal J}_{jk}(\bfx_0,\tau)e^{-i\,\bfk\cendot(\bfx_0-\bfx)} \right\rangle
\label{res-tens}
\eeq
Here, $\bfk = \bfK[(n-1)\tau]$ as defined before. Here, we have $K_i(n\tau)=\gamma_{ji}(-\tau)K_j[(n-1)\tau)]$. Thus, we can note that $K$ at $(n-1)\tau$ is related to the wavevector at $n\tau$ by inverse shearing transformation described in \Eq{shear-Trans}. Note here that, the response tensor ${\cal G}_{ij}$ depends on the time-step $(n-1)\tau$ through $\bfK$, where $\bfK$ is continuously sheared till the time $(n-1)\tau$. Since, we have neglected diffusion term in the induction equation, we will always have the growth of the magnetic field $\bfH(\bfx,t)$ in a single realization
of the ensemble. However, to know the growth at large scale, we have defined the ensemble averaged mean-field $\left\langle\bfH(\bfx,t)\right\rangle$, which may or may not grow. For example, if $h=0$, the mean-field $\left\langle\bfH(\bfx,t)\right\rangle$ will decay (see subsection \ref{average}), whereas $\bfH(\bfx,t)$ might grow at small scales
due to SSD. Here, we are only concerned with the growth of mean magnetic field, or the first moment.

\subsection{Growth rate and cycle periods of the magnetic field}
We can say, a given velocity field will lead to dynamo, if there is a exponential growth of magnetic field in time \citep{DMRS84,Mol84}. In the renovating flow, we are interested in the behaviour of the magnetic field at longer times i.e., as $n\to\infty$. Because, the flow is stationary in the discrete translation of times $n\tau$, $n=1,2,3\dots$, we can consider velocity field as stationary for long times ($n\tau\gg\tau$). Hence, it became possible to construct the eigenvalue problem in any interval ($[(n-1)\tau,n\tau]$) for the evolution of the mean-magnetic field (see \Eq{MatrixEq}) in previous subsection. The magnetic field will grow, if the magnitude of the leading--complex--eigenvalue of the response tensor (given in \Eq{res-tens}) is greater than unity. And the final magnetic field will be the eigenvector of response tensor corresponding to that leading eigenvalue, irrespective of the magnitude and direction of the initial magnetic field. If $\sigma$ is the leading eigenvalue, we can define the exponential ($\exp(\lambda t)$) growing exponent $\lambda$ as, 
\beq
\lambda = \frac{1}{\tau}\ln\sigma(\bfk,\tau) = \frac{1}{\tau}\ln|\sigma|+i\frac{{\rm arg}(\sigma)}{\tau}
\eeq

Since, the response tensor depends on the interval in which magnetic field growth is considered, eigenvalues will also depend on the corresponding interval through $\bfk = \bfK[(n-1)\tau]$. This will lead to the important conclusion i.e., non-axisymmetric modes will decay eventually, which will be elucidated in Section~\ref{nonaxisy}. The real part will define the growth rate of the magnetic field, whereas the imaginary part will define frequency of the wave, which will be used to calculate the cycle period of the dynamo wave. Below, we give expressions for both, 

\beq
\gamma = \frac{1}{\tau}\ln|\sigma|\,,\qquad P_{\rm cyc} = \frac{2\pi\,\tau}{{\rm arg}(\sigma)}
\label{growth_rate}
\eeq
where ${\rm arg}(\sigma)=0$ represent the standing wave and ${\rm arg}(\sigma)\neq 0$ indicate the travelling dynamo wave. 

\subsection{Method of averaging}
\label{average}
To compute the Green's tensor given in \Eq{res-tens}, we need to obtain the Jacobian of transformation between fluid particle at position $\bfx$ at $n\tau$ with the initial position $\bfx_0$ at time $(n-1)\tau$. For general velocity field, we need to solve for $\bfx$ from the equation $\ud \bfx/\ud t = \bfv(\bfx,t)$, which itself is a formidable task. Since, the velocity field considered in \Eq{vel} has constancy of phase ($\bfQ\cendot\bfX=\bfq\cendot\bfx_0$) due to the incompressibility condition ($\bnabla\cendot\bfv=0$), we can integrate the velocity field to obtain the Jacobian of the transformation as

\beq
{\cal J}_{ij} =  \frac{\partial x_i}{\partial x_{0j}} = \delta_{ij}+ q_j\left[ \widetilde{a}_i(t,\bfq)\cos(\bfq\cendot\bfx_0 + \psi) - h\,\widetilde{c}_i(t, \bfq)\sin(\bfq\cendot\bfx_0 + \psi)\right]
\eeq
where 
\begin{align}
\widetilde{\bfa}(t,\bfq)& = \displaystyle\int\limits_{(n-1)\tau}^{n\tau}  [\bfA - S (t-t_0) A_1\ey] \ud t\\
\widetilde{\bfc}(t,\bfq) & = \int\limits_{(n-1)\tau}^{n\tau}  [\bfC - S (t-t_0) C_1\ey] \ud t
\end{align}
Here $t_0=(n-1)\tau$. We can easily average the Green's tensor in \Eq{res-tens} over the phase $\Psi$ (see KSS12 for details) to get,

\beq
G_{ij}(\bsy{k})=\gamma_{ik}(\tau)\left\langle \delta_{kj} J_0\left( \Delta\right)-ih
\frac{q_j\left[\bfk\cross(\widetilde{\bfa}\cross\widetilde{\bfc})\right]_k }{\Delta}
J_1\left( \Delta\right)\right\rangle_{\bfq,\bfa}
\label{shear_turbulence_tensor}
\eeq
where $\Delta= \sqrt{(\bsy{k\cdot\widetilde{a}})^2+h^2(\bsy{k\cdot\widetilde{c}})^2}$ and, $J_0$ and $J_1$ are Bessel functions of order zero and one, respectively. Because, we have $J_0(y)\leq1$, $\forall\, y$, the term that is relevant for the dynamo action would be the second term in \Eq{shear_turbulence_tensor} containing the parameter $h$, which characterizes the kinetic helicity . When $h=0$, i.e., strictly non helical case, there is no mean-field dynamo as already pointed out in KSS12.
We have chosen $h=1$, which corresponds to maximally helical flow, throughout this work.

We perform remaining averages--over ($\bfq$,$\bfa$,$\bfc$)--numerically. We know that, $\bfq$, $\bfa$ and $\bfc$ vectors form orthogonal triad.  Using the three Euler angles for the rigid body rotation, we can relate the triad $(\bfq,\bfa,\bfc)$ to the direction of magnetic field wave vector $(k_1,k_2,k_3)$ at time equal zero.  We use Gauss quadrature methods to perform all the integrals. Further details will be given in the next paper II, which focus on the role of helicity fluctuations on the growth of mean magnetic field.  
We numerically determine the eigenvalues of the response tensor
$\hat{\cal G}$ which governs the evolution of mean magnetic field.
The growth rate and cycle period can then be obtained from Eq.~\ref{growth_rate}.
Below we discuss the non-axisymmetric and axisymmetric
mean field dynamos.
We present our findings in the non-dimensional units, henceforth. All the quantities are suitably normalized with respect to eddy turn-over time of the flow at the beginning of each interval, $T = 1/qa$, and the wave vector $q$.
Quantities with over-tilde are made dimensionless in this way, e.g.,
$\widetilde{\bfk}=\bfk/q$, $\widetilde{\gamma}=\gamma T$,
$\widetilde{\alpha}_{ij}=\alpha_{ij}/a$, and so on.

\section{Decay of non-axisymmetric modes of mean magnetic field}
\label{nonaxisy}

\begin{figure*}
\includegraphics[width=\columnwidth]{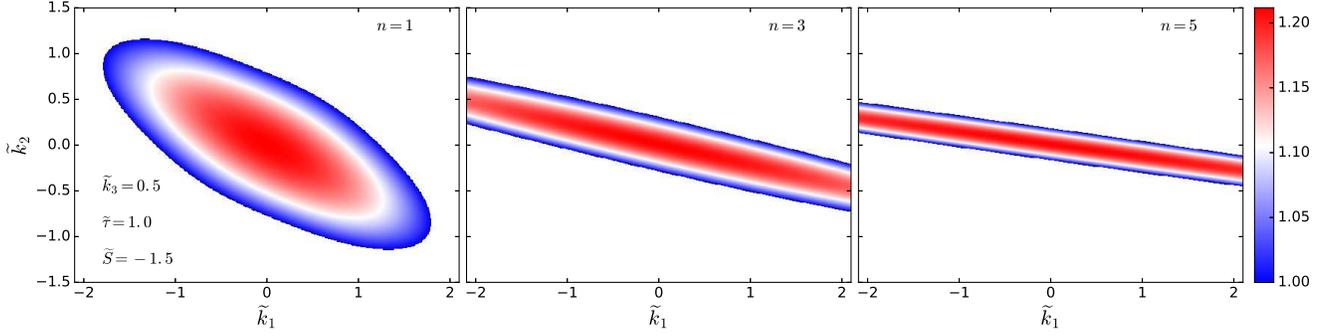}
\caption{Contours of the magnitude of leading eigenvalue
$\sigma$ in $k_1-k_2$ at $k_3=0.5$, obtained at
the end of first (left), third (middle) and fifth (right) interval $n$;
time increases from left to right as $n=t/\tau$.}
\label{nonaxi}
\end{figure*}

We show in this subsection that the non-axisymmetric mode of the mean-magnetic field decays asymptotically. Since shearing flows are anisotropic in all three directions, the eigenvalues of the response tensor for the magnetic field will also be anisotropic in the directions of $k_1$, $k_2$ and $k_3$. 
 We have decomposed the magnetic field in the shearing waves (see \Eq{Fou-Trans2}), where we have used time dependent wave vector. If $(k_1,k_2,k_3)$ be the wave vector of the magnetic field at time $t=0$, then at $t=n\tau$, it would become $(k_1-n\,S\tau\,k_2,k_2,k_3)$. 
 
 Let us denote the magnetic field by the column vector $\hat{H}$ and response tensor by the square matrix $\hat{{\cal G}}$ (see \Eq{MatrixEq}). Let $\hat{H}_0$ be the initial magnetic field at $t=0$, then the magnetic field at $t=n\tau$ is given by
\beq
\Hat{H}_n = \hat{\cal G}_n\dotsc\hat{\cal G}_2\hat{\cal G}_1\hat{H}_0
\eeq 
 where $\hat{\cal G}_n$ indicate the response tensor in $n^{\rm th}$ interval with sheared wavevector $(k_1-n\,S\tau\,k_2,k_2,k_3)$.

We now describe the procedure to determine $\Hat{H}_n$ iteratively.
At the end of the first interval, we get
$\Hat{H}_1=\hat{\cal G}_1 \Hat{H}_0$.
We find that one of the eigenvectors of $\hat{\cal G}_1$ does not satisfy $\bfk\cendot\bfH_0=0$. Let $\hat{V}_{01}$ and $\hat{V}_{02}$ be the eigenvectors (with corresponding eigenvalues being $\sigma_{01}$ and $\sigma_{02}$), such that, they are orthogonal to the direction of $\bfk$.  We can now express the initial magnetic field in terms of the eigenvectors as 
\beq
\hat{H}_0 = C_{01} \hat{V}_{01}+C_{02}\hat{V}_{02}
\eeq
where $C_{01}$ and $C_{02}$ are some complex constants.
The quantity $\Hat{H}_1$ thus becomes:
\beq
\hat{H}_1 = C_{01}\,\sigma_{01} \hat{V}_{01}+C_{02}\,\sigma_{02}\hat{V}_{02}
\label{h1}
\eeq

 For the next iteration, that is at $t=2\tau$, we have $\hat{H}_2=\hat{\cal G}_2 \Hat{H}_1$. The response tensor $\hat{\cal G}_2$ 
is modified because the magnetic field wave vector $(k_1,k_2,k_3)$ would shear to $(k_1-S\tau\,k_2,k_2,k_3)$. Hence, $\hat{V}_{01}$ and $\hat{V}_{02}$ will not anymore be the eigenvectors of $\hat{\cal G}_2$. We need to express the eigenvectors $\hat{V}_{01}$ and $\hat{V}_{02}$ in terms of the eigenvectors of $\hat{\cal G}_2$. Similarly, let $\hat{V}_{11}$ and $\hat{V}_{12}$ be the eigenvectors (with corresponding eigenvalues being $\sigma_{11}$ and $\sigma_{12}$) of the response tensor $\hat{\cal G}_2$ orthogonal to the direction of $(k_1-S\tau\,k_2,k_2,k_3)$. Then we have,
\beq
\hat{V}_{01} = C_{11}^{(1)}\hat{V}_{11}+C_{12}^{(1)}\hat{V}_{12} \qquad \hat{V}_{02} = C_{11}^{(2)}\hat{V}_{11}+C_{12}^{(2)}\hat{V}_{12}
\label{secInt}
\eeq 
where $C_{11}^{(1)}$, $C_{12}^{(1)}$, $C_{11}^{(2)}$ and $C_{12}^{(2)}$ are again some complex constants.\footnote{Note here that, we have suppressed the third component of $\hat{V}_{01}$ and $\hat{V}_{02}$. Because at every interval we need to satisfy the solenoidality condition ($\bnabla\cendot\bfH$=0).} Using \Eq{secInt} in \Eq{h1}, we get
\beq
\hat{H}_2 = (\dots\sigma_{01}\sigma_{11}+\dots\sigma_{02}\sigma_{11})\hat{V}_{11} + (\dots\sigma_{01}\sigma_{12}+\dots\sigma_{02}\sigma_{12})\hat{V}_{12}
\eeq 
 where ellipsis indicate pre-multiplied factors, such as multiplication of complex coefficients. As we continue to iterate in the above manner by expressing the preceding eigenvectors in terms of the current eigenvectors (like in \Eq{secInt}), we get $2^{n+1}$ terms for the magnetic field $\hat{H}_n$ at the end of $n^{\rm th}$ interval;
at every interval, the number of terms are doubled. Of the two relevant eigenvectors at any interval, let us say that we have, $|\sigma_{n1}|>|\sigma_{n2}|$, then in those $2^{n+1}$ terms, there will be a term of the kind $\sigma_{01}\sigma_{11}\sigma_{21}\dots\sigma_{n1}$ whose magnitude will be the largest compared to other terms. To demonstrate that non-axisymmetric modes decay eventually, it is enough to show that this term decays after some interval of time. 
 
\begin{figure}
    \centering
    \hspace{-0.4cm}
    \includegraphics[width=0.5\columnwidth,height=0.37\columnwidth]{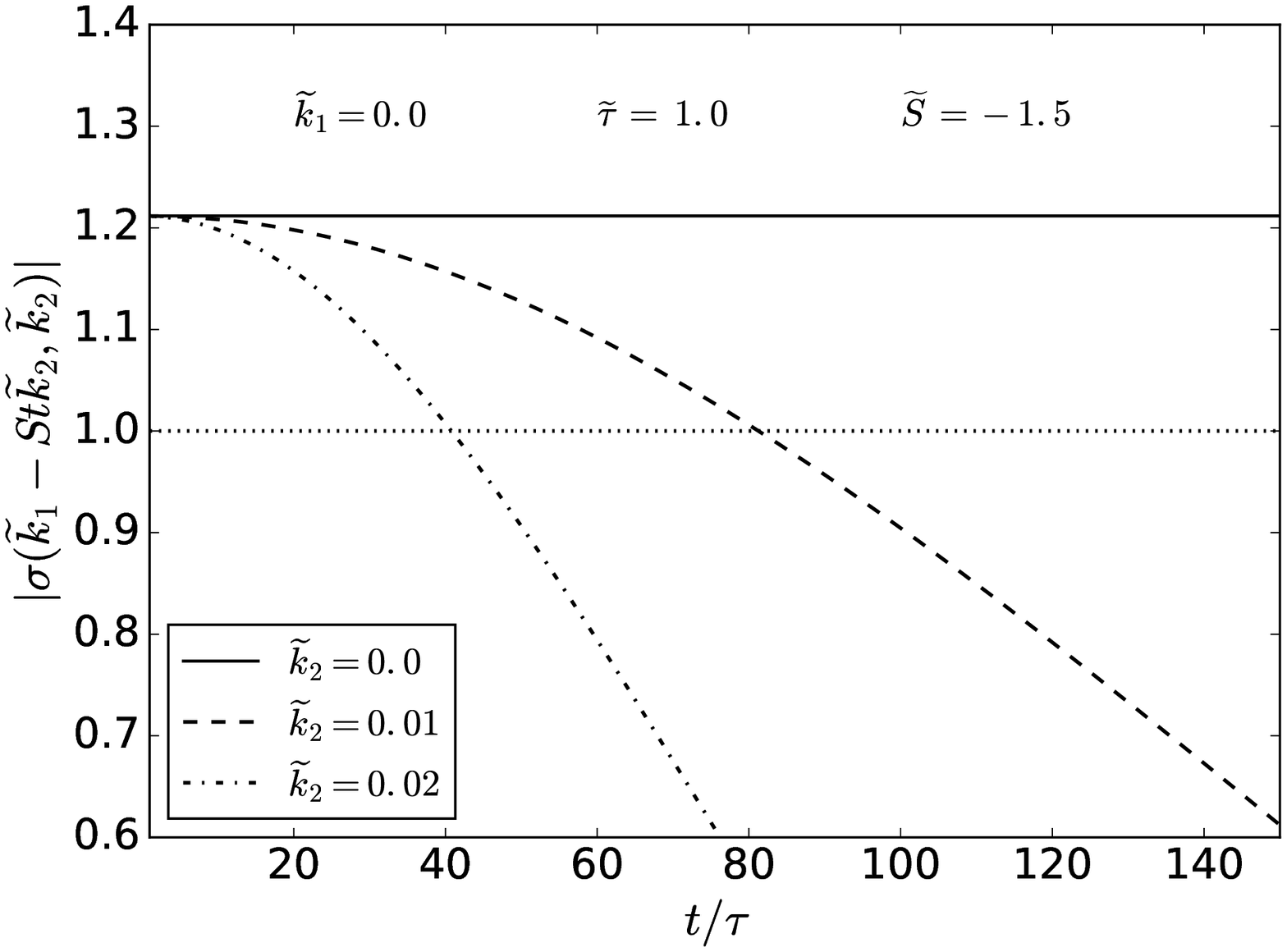}
    \includegraphics[width=0.5\columnwidth,height=0.37\columnwidth]{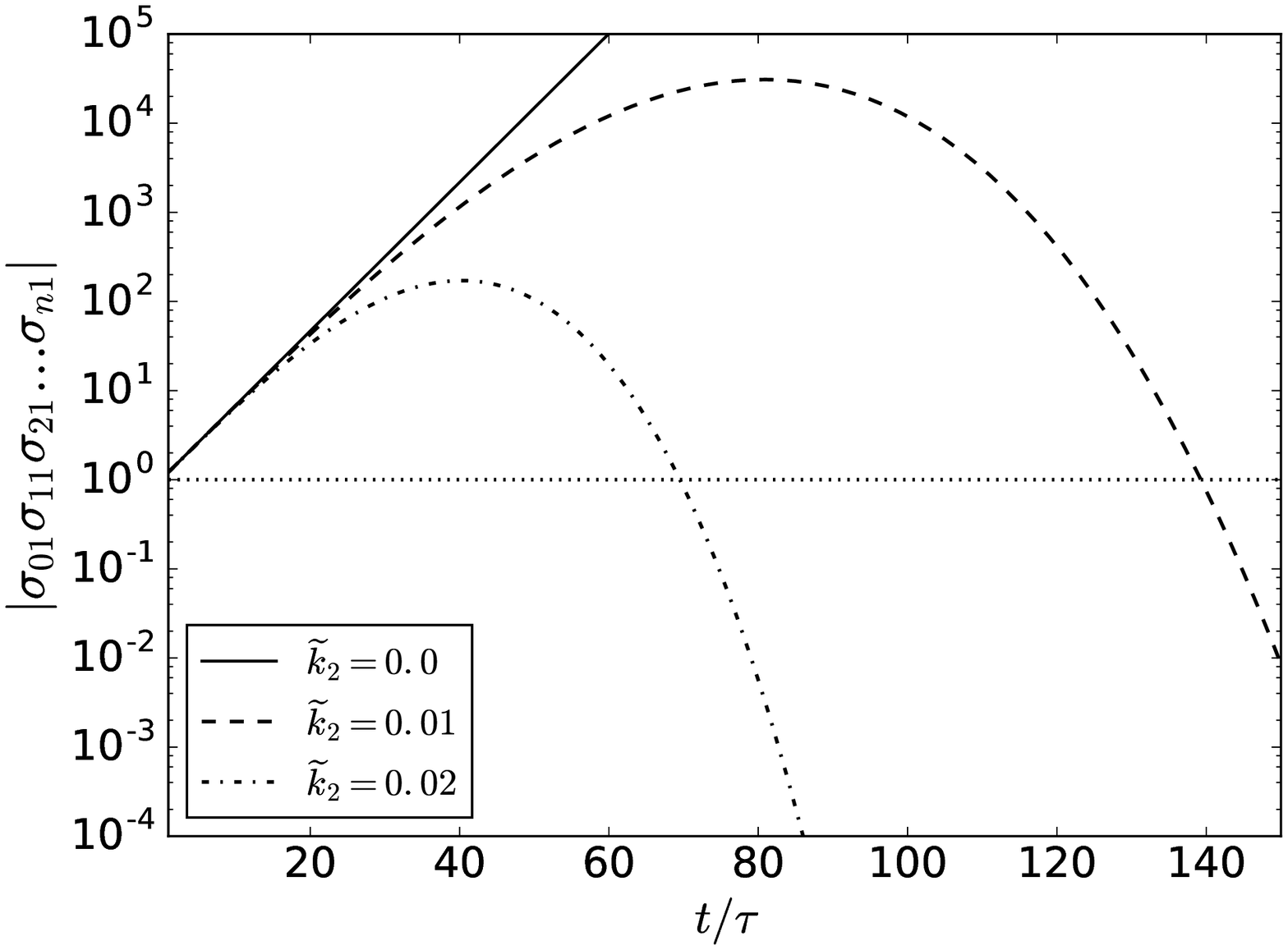}
    \caption{(Left) Magnitude of the leading eigenvalue and (right)
    cumulative product of the maximum eigenvalues as a function of
    $n=t/\tau$, i.e., the interval of the renovating flow.}
    \label{cumsum}
\end{figure}

In \Fig{nonaxi} we have shown the magnitude of the largest eigenvalues $\sigma_{n1}$ in the $k_1-k_2$ plane for $k_3=0.5$ at the end of first ($t=\tau$), third ($t=3\tau$), and fifth ($t=5\tau$) interval. We have highlighted the area where $|\sigma_{n1}|>1$, which represents the
transient growth region. As time is increasing, the wave vector is continuously sheared in the $k_1$ direction, the transient growth region is stretched in the $k_1$ direction, and it diminishes in the $k_2$ direction. As time continues, the growth region aligns with the
$k_2=0$ axis, and eventually vanishes. To make this point clear, let us consider the wavevectors, which lie close to maxima of the contours shown in \Fig{nonaxi}, i.e., (0,0) in $k_1-k_2$ plane. In the left panel of \Fig{cumsum}, we have shown the largest eigenvalue as the function of intervals for three different values of $k_2$. When $k_2=0$ (axisymmetric mode), the magnitude of the largest eigenvalue remains constant as the wavevector $(k_1,0,k_3)$ remains same across the intervals. Therefore the cumulative product of the magnitude of the eigenvalues $|\sigma_{01}\sigma_{11}\sigma_{21}\dots\sigma_{n1}| = |\sigma_{01}|^n$, increases monotonically leading to the exponential growth of the axisymmetric mode of the magnetic field (see \Fig{cumsum}: right panel, black solid line). For $k_2\neq 0$, the wave vectors are time dependent. As time increases, the value of $k_1-St\,k_2$ increases (for negative $S$), eventually the wave vector moves out of the transient growth region i.e., the region where $|\sigma_{n1}|>1$ (see \Fig{nonaxi}). As shown in \Fig{cumsum} (left panel), for $k_2=0.01$ ($k_2=0.02$), the magnitude of eigenvalue falls below unity around $t=80\tau$ ($t=40\tau$). The magnitude of the cumulative product of the largest eigenvalues $|\sigma_{01}\sigma_{11}\sigma_{21}\dots\sigma_{n1}|$ increases for some time (see right panel in \Fig{cumsum}) and then it starts to decrease, and falls below unity leading to the decay of the mode. Hence, non-axisymmetric modes will only have transient growth before decaying eventually. Even though, the analysis of this section is made with the particular velocity field, it's validity remains general. Because, essential argument to show the asymptotic decay needs only two ingredients: the magnetic wave vector is time-dependent, which is a consequence of background shear flow; and the growth region is limited in the $k$-space, which is the consequence of the finite correlation of velocity field rather than it's particular choice. Therefore, in the kinematic dynamo regime, non-axisymmetric mode have no active role to play. 
  
\section{Growth of axisymmetric modes of mean magnetic field}
\label{axi}

\begin{figure*}
\includegraphics[width=\columnwidth]{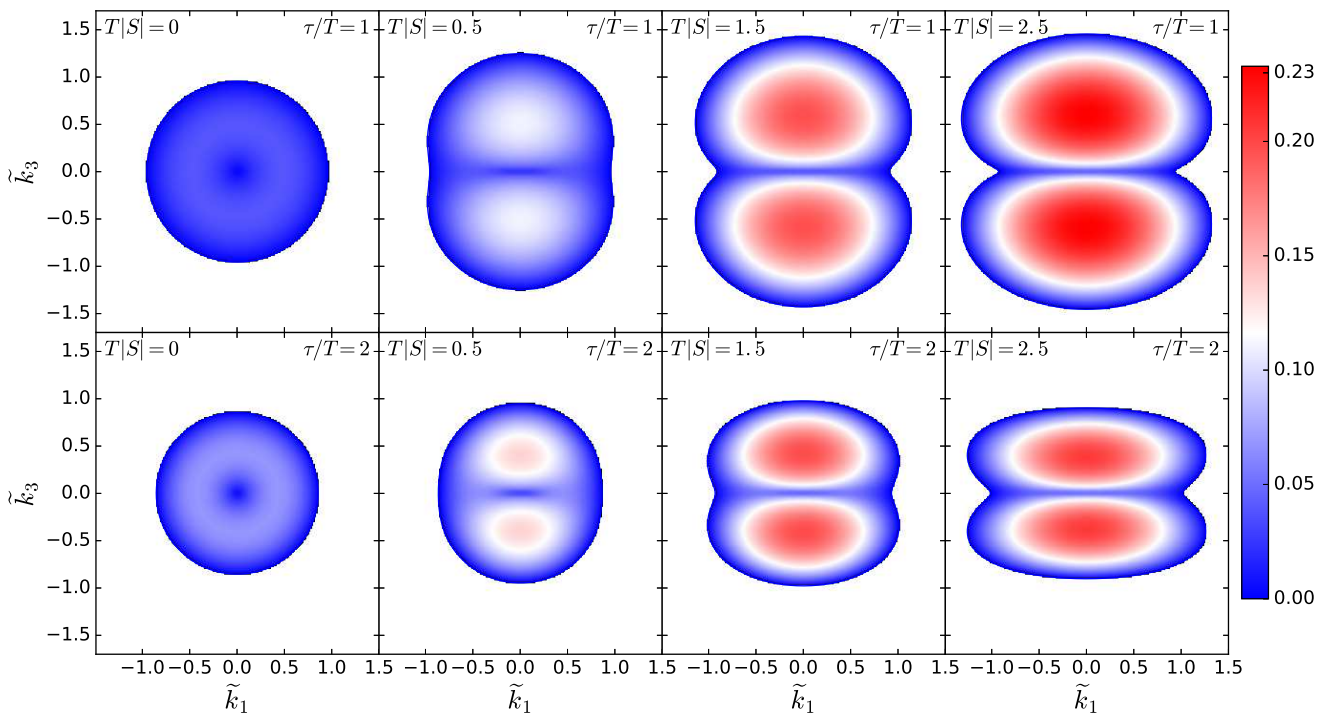}
\caption{Contours of growth rate $\tilde{\gamma}\geq0$
in $k_1-k_3$ plane for axisymmetric (i.e., $k_2=0$) mean
field dynamos. Shear increases from left to right,
with $\tau/T=1$ (top) or $\tau/T=2$ (bottom). Regions outside
the outermost (blue) contour do not support a dynamo instability.}
\label{k1k3_t1t2_v2}
\end{figure*}

From now onward we focus only on the axisymmetric
solutions for which $k_2=0$, as the non-axisymmetric
modes are expected to decay as discussed above.
Since, for axisymmetric modes the 
eigenvalues and eigenvectors are constant in time, we just need
to consider it's growth rate in a single interval
(see the black solid line in left panel of \Fig{cumsum}).
With $\sigma$ being the leading eigenvalue of the response tensor,
we find that the magnetic field at the end of the $n^{\rm th}$ interval
is given by $H_n = \sigma^n H_0$ where $H_0$ is the initial magnetic field,
also assumed to be the corresponding eigenvector.
Figure~\ref{k1k3_t1t2_v2} shows contours of normalized growth rate $\widetilde{\gamma}=\gamma\, T$, with its positive
values indicating exponentially growing solutions in
$k_1$-$k_3$ plane for axisymmetric mean field
dynamos, as functions of the two parameters,
the shear rate $S$, and the renovation time $\tau$.
Note that $T=1/qa$ is the eddy turnover time
of the random helical flow at the beginning of each interval.
Regions enclosed within the outermost (blue) contours
in \Fig{k1k3_t1t2_v2} is referred as
dynamo regions.

For zero shear, dynamo regions are circularly symmetric about
the origin $k_1=k_3=0$; see leftmost panels in
\Fig{k1k3_t1t2_v2}. The other panels there reveal that
the dynamo regions gets bifurcated for non-zero shear.
Note that the maximum growth occurs along $k_3$ axis when
$k_1=0$. Therefore, without any loss of generality,
and in order to capture the branch containing the fastest growing
mode, we set $k_1=0$ henceforth. We also find from this figure
that the growth rate is symmetric about the point $k_3=0$, and
therefore we consider only positive values of $k_3$ to
explore its behaviour as a function of wavenumber.
Thus, we have now set $k_1=k_2=0$, which is
equivalent of taking average over entire $X_1$-$X_2$ plane,
i.e., the plane of background shear, and we study one dimensional
mean field dynamo modes propagating along $X_3$ direction.

Interestingly, the wavenumber corresponding to the
fastest growing mode, denoted by $k_\ast$, varies
non-monotonically with the strength of shear $|S|$ when
the renovation time $\tau$, which is the same
as the correlation time of the flow, becomes comparable to the
eddy turnover time $T$. This is better shown in
\Fig{k3max_wS}, where various curves correspond
to different choices of $\tau/T$. However,
when $\tau/T \ll 1$, i.e., when the memory effects are
unimportant as the random flow is nearly of white-noise type,
the maximum growth occurs at progressively smaller spatial
scales ($\sim k_\ast^{-1}$) with increasing shear strengths;
sufficiently strong shear produces magnetic field preferentially
at scales smaller than the eddy size given by $q^{-1}$,
i.e., $k_*/q > 1$, as may be seen from the dash dotdotted curve.
It is only when $\tau/T$ becomes of order unity, the shear
gets some time to act and it changes the scenario even
qualitatively; see also \cite{SS09}. It promotes
a genuine large-scale dynamo as $k_*/q < 1$ for a whole
range of shear, i.e., in this case, the mean magnetic
field grows maximally at scales larger than the eddy scale.
Also note that at fixed shear, $k_*$ systematically decreases
when $\tau$ increases; see also Appendix (A) where we show a
comparison with a case when non-shearing waves are used to model
the renovating flow.

\subsection{Growth rate}

\begin{figure}
\centering
\includegraphics[width=0.7\columnwidth]{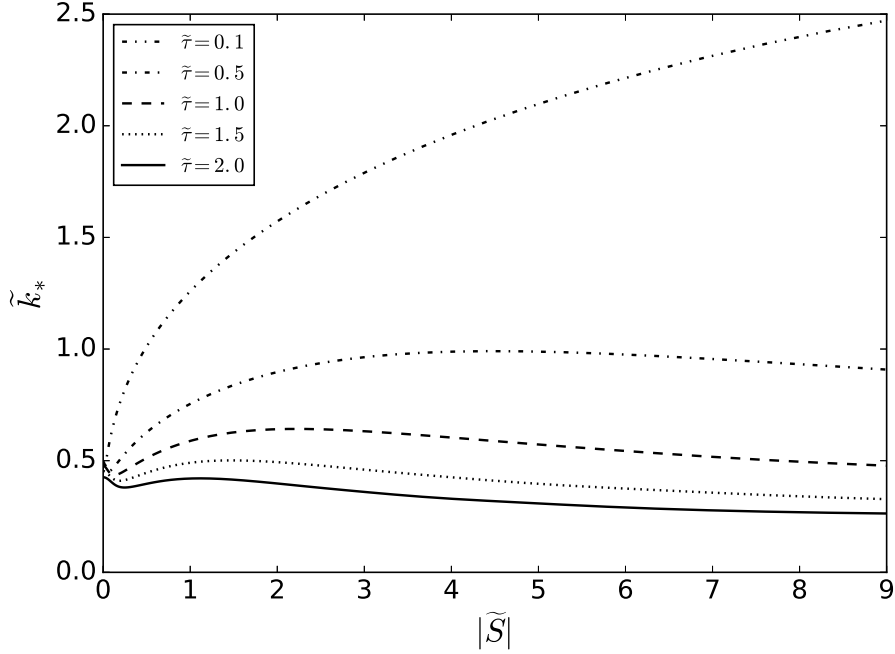}
\caption{Dependence of $k_\ast/q$ (wavenumber corresponding
to the fastest growing mode) on shear for different choices
of $\tau/T$.}
\label{k3max_wS}
\end{figure}

\begin{figure}
\includegraphics[width=0.5\columnwidth]{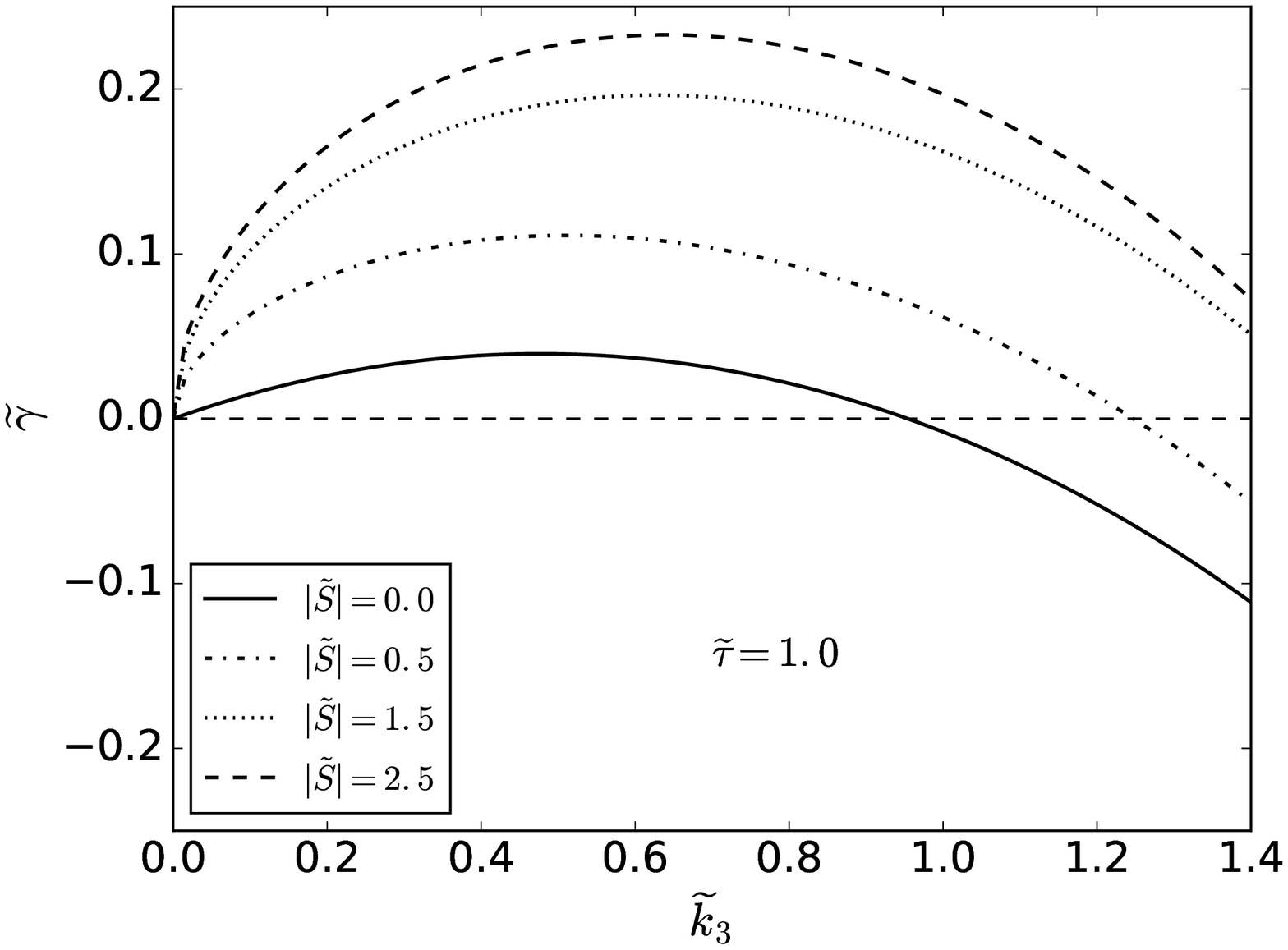}
\includegraphics[width=0.5\columnwidth]{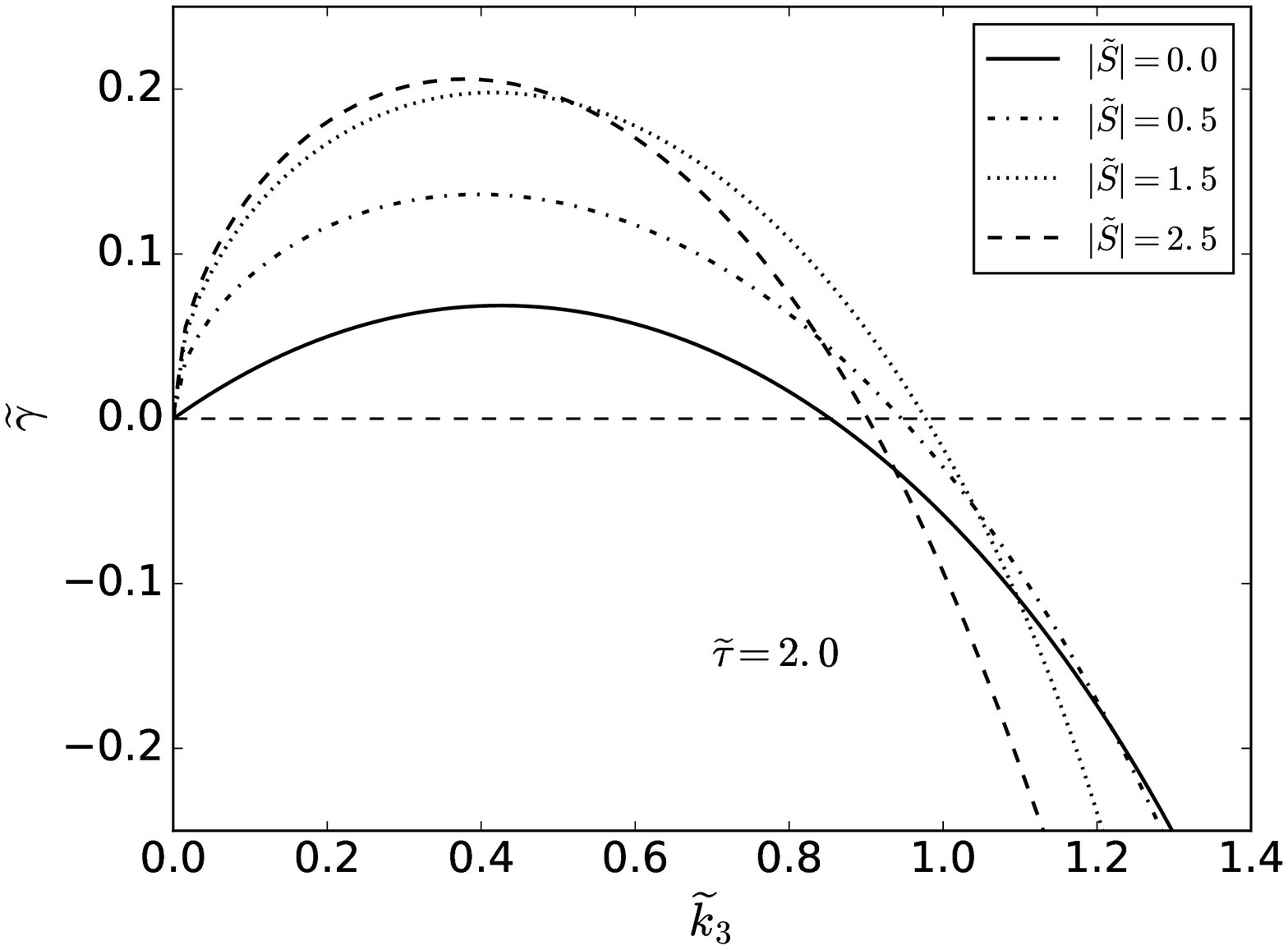}
\caption{Normalized growth rate $\tilde{\gamma}$ as a function of wavenumber $k_3$,
for $\tau/T=1$ (left) and $\tau/T=2$ (right). Different curves
in each panel correspond to different values of shear rate $S$.}
\label{Relam_wk3}
\end{figure}

In \Fig{Relam_wk3} we show the behaviour of normalized growth
rate $\widetilde{\gamma}$ of the mean magnetic field as a
function of wavenumber $k_3$. We have chosen two large values
for the correlation time $\tau$ in the two panels, where
different curves in each panel correspond to different values of
shear rate $S$; $\tau/T=1$ and $2$ in left and right panels,
respectively. Regardless of the strength of the shear,
including its zero value,
the growth rate first increases from zero as
a function of $k_3$, attains a maximum, then it decreases
to become negative at sufficiently large wavenumbers.
Note again that the maximum lies at wavenumbers that are
smaller than the one corresponding to random eddies, and
magnetic fields at sufficiently large wavenumbers are
always suppressed.
Looking first at the more reasonable case with $\tau/T=1$,
we find that the growth rate increases at all the wavenumbers
shown, when the shear parameter $\widetilde{S}$ is increased
from zero to a moderately large values; see left panel
of \Fig{Relam_wk3} where $0\leq \widetilde{S}\leq 2.5$.
However, the behaviour is more complicated when $\tau/T=2$,
as, at fixed $k_3\gtrsim q$, shear leads to suppression
of mean magnetic fields; see the right panel.
Nevertheless, the peak of the growth rate remains
at much smaller wavenumbers, thus enabling a large-scale dynamo.

\begin{figure}
\centering
\includegraphics[width=0.7\columnwidth]{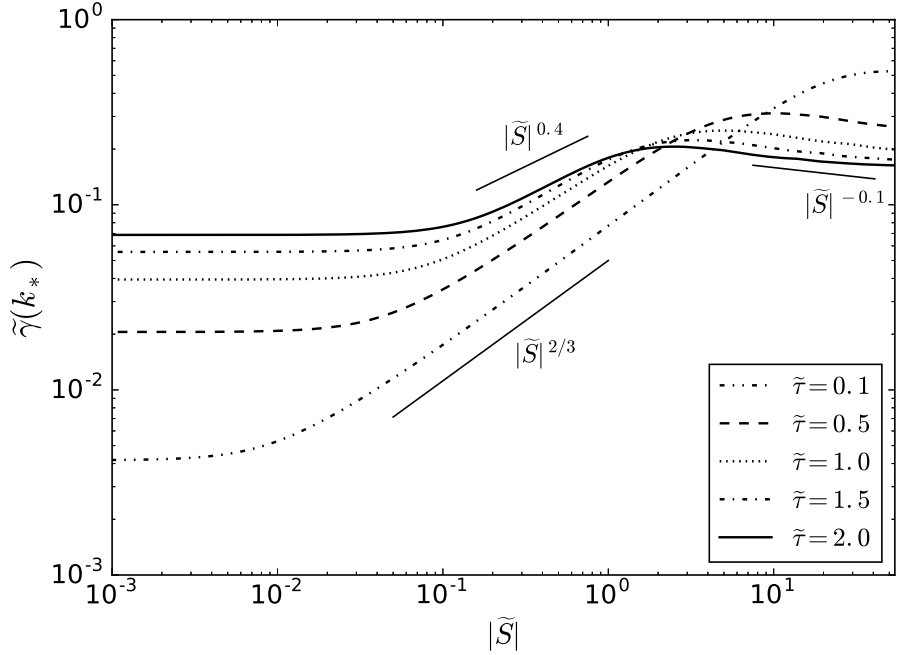}
\caption{Growth rate of fastest growing mode as a function of shear
for different choices of velocity correlations times $\tau$.}
\label{Relam_k3max_wSlog}
\end{figure}

Now we turn to the dependence of the growth rate of
mean-field dynamo on the shear. We saw earlier in \Fig{k3max_wS}
that the wavenumber ($k_\ast$) corresponding to the fastest
growing mode is itself a function of shear.
Therefore, in \Fig{Relam_k3max_wSlog}, we show the shear
dependence of the growth rate $\widetilde{\gamma}$ at $k_\ast$.
Remarkably, the growth rate shows a non-monotonic trend
with the shear strength in more realistic regime when
the correlation time $\tau$ of the random helical
flow becomes comparable to the eddy turn over time $T$;
see, e.g., dotted or solid curves in \Fig{Relam_k3max_wSlog}.
The growth rate here is always positive, even when
shear approaches zero, due to the fact that we have chosen
maximally helical random flow ($h=1$).
Initially, the growth rate remains constant as shear
increases, and this constant is a function of correlation time.
For $\widetilde{\tau}=0.1$, we have $\widetilde{\gamma}\simeq 4\times 10^{-3}$ and for $\widetilde{\tau}=2.0$, we have $\widetilde{\gamma}\simeq 6,5\times 10^{-2}$ (see \Fig{Relam_k3max_wSlog}); dynamo becomes more efficient at
a fixed weak shear as correlation time increases.
As shear strength increases, the growth rates starts to increase, as can be seen from \Fig{Relam_k3max_wSlog}. The growth rate varies as $\widetilde{\gamma}\sim S^{2/3}$ for $\widetilde{\tau}=0.1$  and as $\widetilde{\gamma}\sim S^{0.4}$ for $\widetilde{\tau}=2.0$, in the intermediate shear range. As the shear strength further increases, the growth rate starts to decrease as $\widetilde{\gamma}\sim S^{-0.1}$ after reaching the maximum. This decrease is happening at large shear because, the transport coefficient like $\alpha_{ij}$ is a function of shear and decreases as the shear strength increases (see \Fig{alpha})\footnote{See also appendix (\ref{unshear_comp}) for the behaviours of $\gamma$ and $k_\ast$, for the case
when flow amplitudes are constant, i.e.,
independent of shear. In such a scenario, transport coefficients also become independent of shear.}. This results in quenching of the dynamo at strong
shear. Our results are in agreement with the work
of \cite{LK08} who also reported dynamo quenching due to
strong shear. Note that this is unlike more popular expectation
based on standard kinematic $\alpha \Omega$ dynamos, where
the dynamo efficiency increases monotonically with shear.
Such expectations have resulted in common notion that the
regions with strongest shear in a system, e.g. the
tachocline in case of the Sun, must be the best reservoirs
of magnetic fields.
We envisage that the dynamo quenching being reported here
in the strong shear regime will be helpful for a
better understanding in this direction.

\subsection{$\alpha$-effect}

By adapting to a standard textbook definition of $\alpha_{ij}$,
we make an attempt to determine the
components of $\alpha$ tensor based on the random
velocity fields we have chosen in our model.
This may provide useful insights for key mechanisms that
govern the properties of the large-scale dynamo
action we study in this work. More precisely, it may help us understand
the reason for dynamo quenching at large shear as shown earlier in
\Fig{Relam_k3max_wSlog}.
We have chosen the following definition \citep[see,][section 7.10]{Mof78}:

\beq
\alpha_{ij}(\tau) = \frac{1}{\tau}\int_0^\tau {\rm d}t
\int_0^t{\rm d}t' \hat{\alpha}_{ij}(t,t') \quad
\mbox{with} \quad
\hat{\alpha}_{ij}(t,t') = \epsilon_{ilk}
\left\langle v_l(\bfx_0,t)\frac{\partial v_k(\bfx_0, t')}{\partial x_j}\right\rangle\,.
\label{alpha_def}
\eeq

\begin{figure}
\includegraphics[width=0.5\columnwidth]{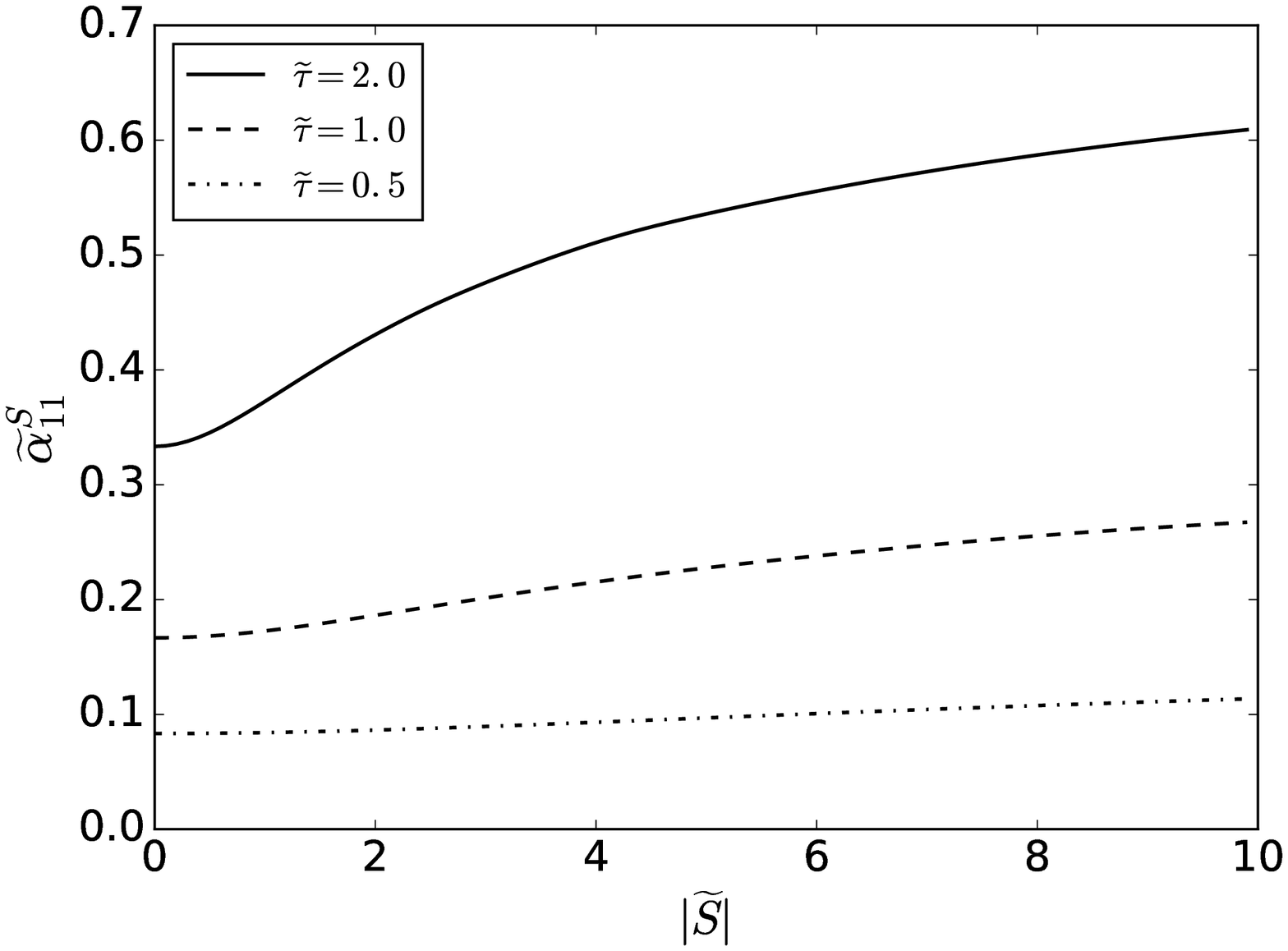}
\includegraphics[width=0.5\columnwidth]{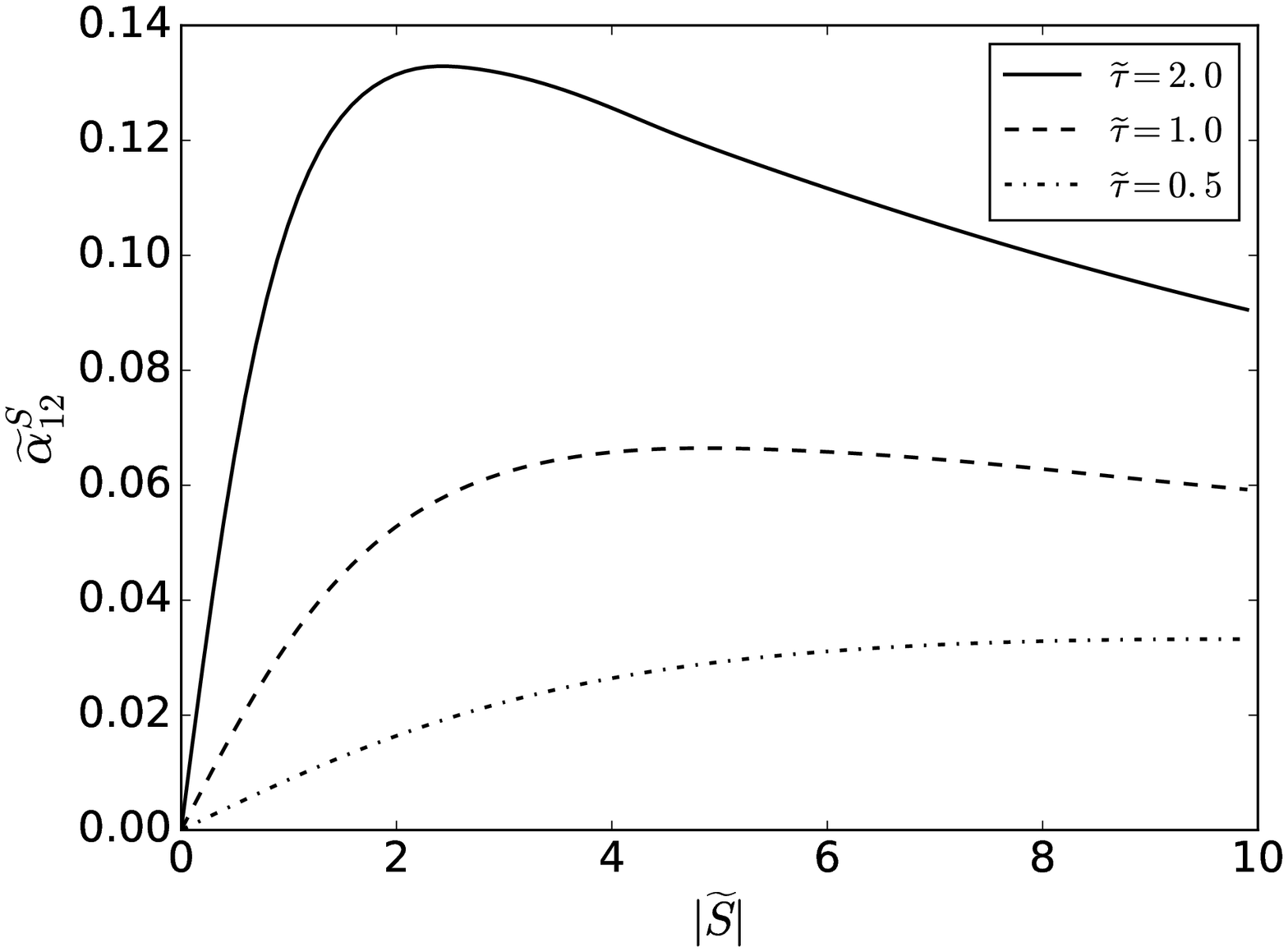}
\includegraphics[width=0.5\columnwidth]{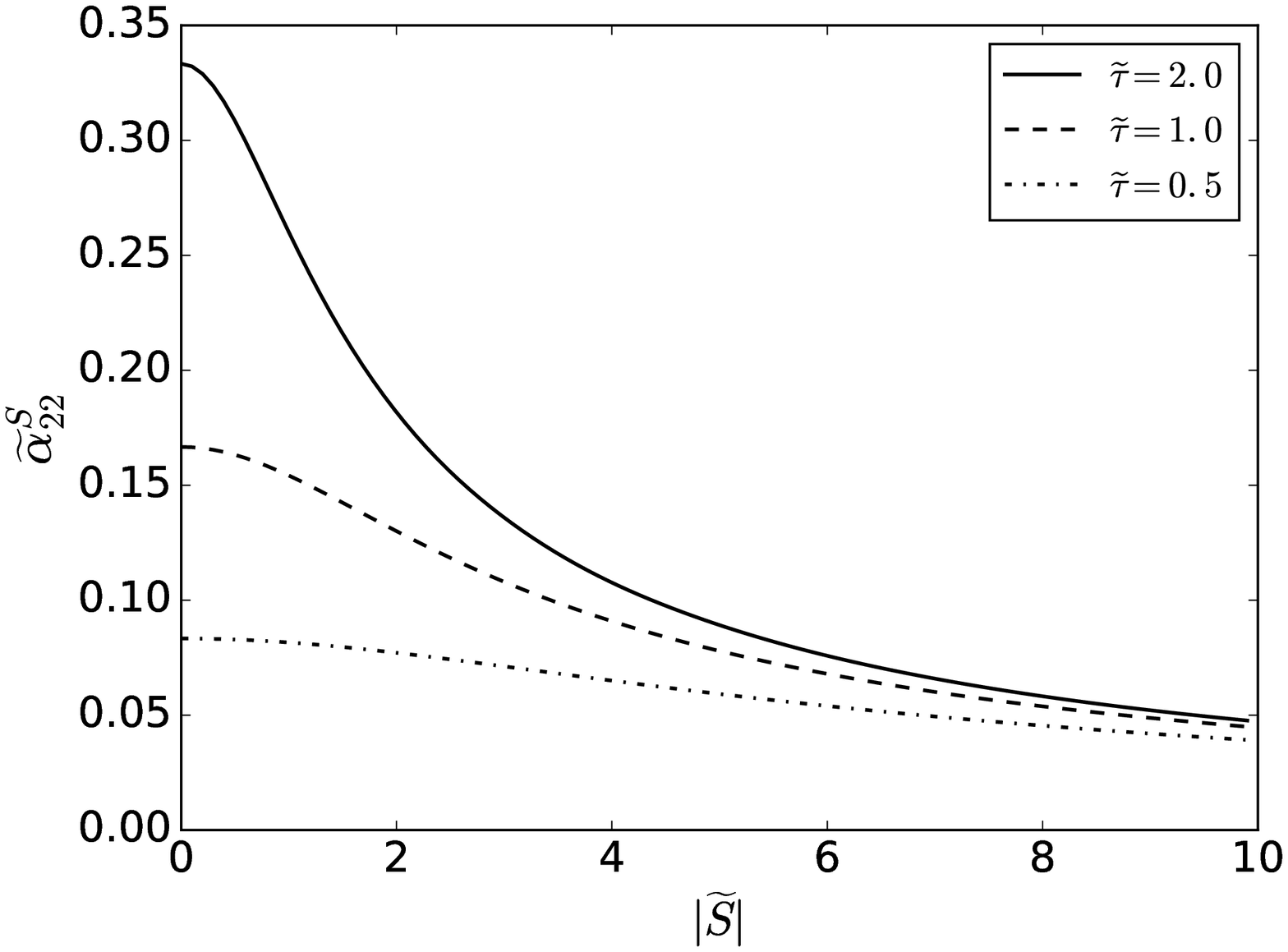}
\includegraphics[width=0.5\columnwidth]{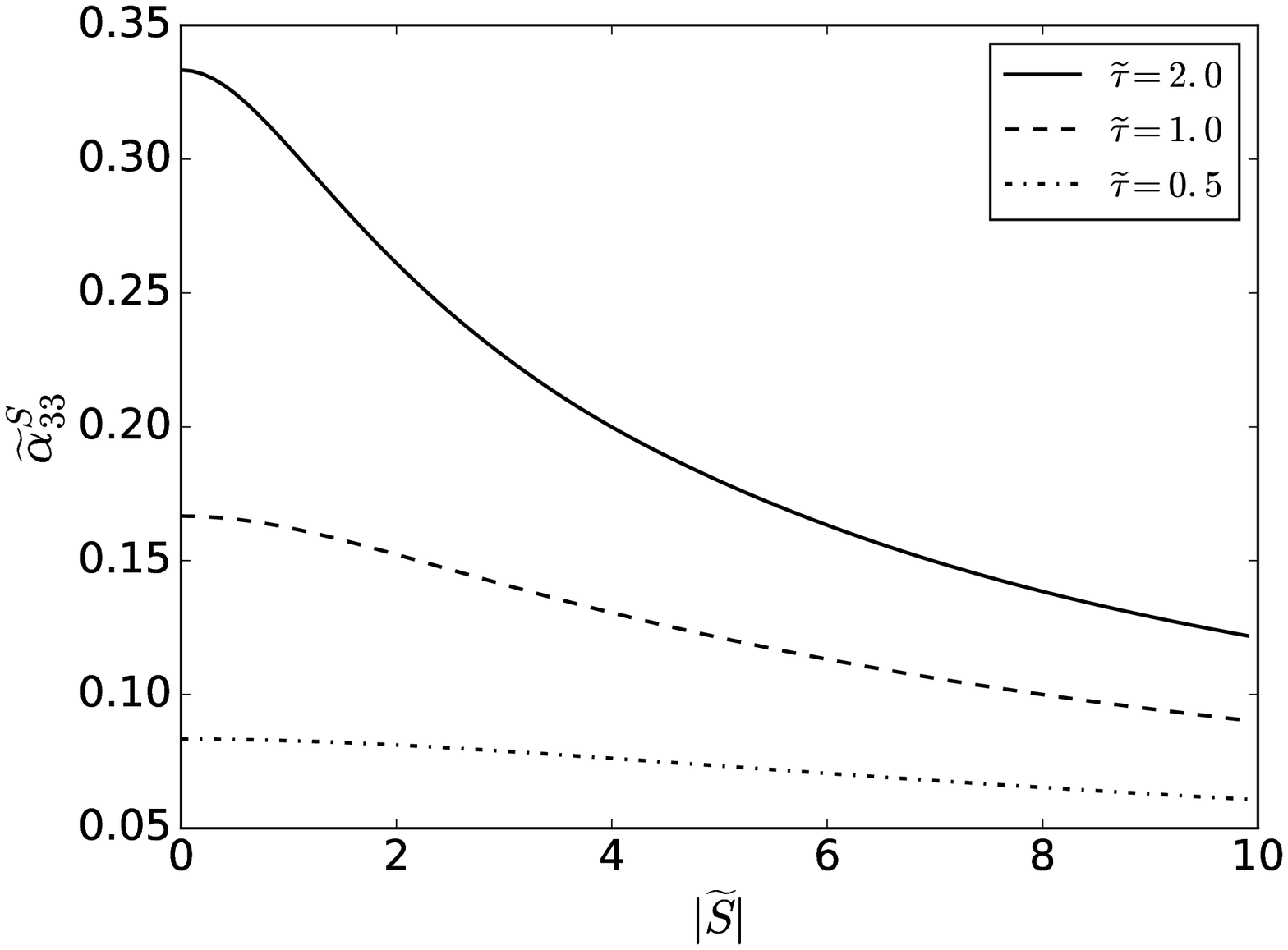}
\caption{Behaviour of non-zero components of symmetric tensor
$\widetilde{\alpha}_{ij}^S$ with $|\widetilde{S}|$. Different curves
correspond to different choices of $\widetilde{\tau}$.}
\label{alpha}
\end{figure}

Here the velocity field $\bfv$ is given from \Eq{vel} in terms of
time-dependent shearing waves with fixed helicity \citep{SS17}.
This evolves for the renovation time interval $\tau$ which
represents one single realization. The average is then taken
over many such realizations, or equivalently, over time
$t=n \tau$ with $n\rightarrow \infty$. The second integral is taken to 
average the $\alpha$-tensor over the interval from $0$ to $\tau$.  Note that while the
velocity $\bfv$ gets randomized after every $\tau$, the
kinetic helicity associated with it stays constant for all times
in the present work. Following \cite{RKR03} we symmetrize the $\alpha_{ij}$
defined in \Eq{alpha_def} as $\alpha_{ij}^S=(\alpha_{ij}+\alpha_{ji})/2$;
the antisymmetric part corresponds to the turbulent pumping which we ignore
in this study. In \Fig{alpha} we show the behaviour of non-zero components
of dimensionless quantity $\tilde{\alpha}_{ij}^S$ as a function of shear;
other components vanish identically for our choice of random velocity field.
While $\alpha_{11}$
shows an increase with $|S|$, $\alpha_{22}$ and $\alpha_{33}$ are
significantly quenched when shear becomes too large. The behaviour of
$\tilde{\alpha}_{12}^S$ is more involved as may be seen from the \Fig{alpha}. The trace of $\alpha$-tensor i.e., $\alpha_{11}+\alpha_{22}+\alpha_{33}$ is also a decreasing function of shear.  Note that the helicity as defined by 
\Eq{hel} is independent of shear rate $S$. 
Also, at a fixed value of shear, magnitudes of all the components increase
with $\tilde{\tau}=\tau/T$, where we note again that $\tau$ and $T$ represent
velocity correlation and eddy turn over times, respectively.
Thus, we find that the $\alpha$-tensor is strongly affected by the presence of shear,
with effect being more pronounced when velocity correlation times $\tau$
are comparable to the eddy turn over time $T$. 

\begin{figure}
\centering
\includegraphics[width=0.8\columnwidth]{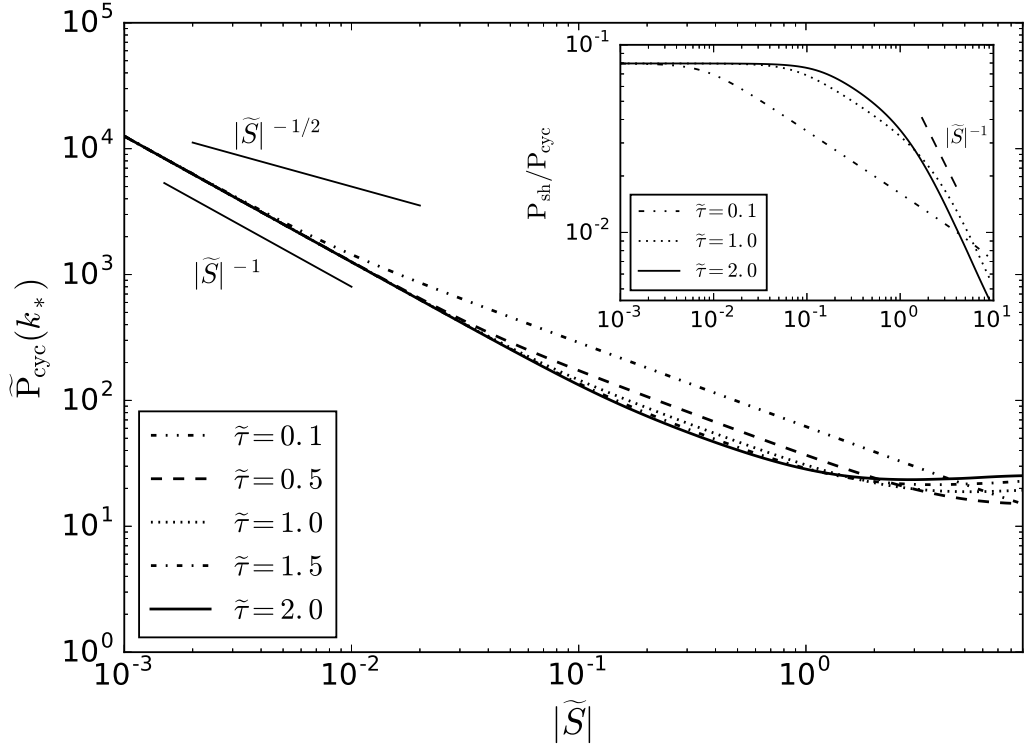}
\caption{Cycle period of dynamo wave as a function of shear for different
values of $\tau$. Inset: Dimensionless quantity $P_{\rm sh}/P_{\rm cyc}$
as a function of shear, where $P_{\rm sh}=1/|S|$ .}
\label{Pcyc_k3max_wS}
\end{figure}

\subsection{Cycle period of dynamo waves}

Another important quantity is the cycle period of the
growing dynamo wave. This is denoted by $P_{\rm cyc}$ and
defined in \Eq{growth_rate}. Its behaviour at $k_\ast$
as a function of shear is shown in \Fig{Pcyc_k3max_wS}.
It falls with shear as $|S|^{-1}$ for all $\tau$ when
shear is weak, but this scaling becomes shallower at larger
values of shear rate. Interestingly enough, $P_{\rm cyc}$
becomes nearly independent of shear, when
normalized absolute shear $|\tilde{S}|\gtrsim 1$
and $\tau/T\sim {\cal O}(1)$.
Thus our present model yields,
for a dimensionless quantity $1/(|S| P_{\rm cyc})$ (see the inset of \Fig{Pcyc_k3max_wS}),
a scaling of (i) $|S|^0$, i.e., independent of shear,
at weak shear, and (ii) $|S|^{-1}$ when shear becomes sufficiently
strong.

Note that the standard
$\alpha \Omega$ ($\alpha^2 \Omega$) dynamo predicts a uniform scaling,
$P_{\rm cyc} \sim |S|^{-1/2}$ ($|S|^{-1}$), with shear.
It is intriguing to note here that
\cite{Ols18}, based on their observational analysis,
found evidence of two distinct population
of stars, inactive and active, which reveal different scalings
in a stellar magnetic activity-rotation diagram.
Standard dynamo models fail to explain the existence of
these branches in such a diagnostic diagram, which
provides a sufficient motivation for further work on
this topic of turbulent dynamo action due to
helicity and shear; see \cite{Ols18}, and
references therein, to appreciate the importance of
studying the cycle periods of dynamos, as this offers a
unique opportunity to test model predictions.
We believe that the new scaling laws which we find in this work
will have implications for the
interpretation of the observations of magnetic activity
cycles seen in \cite{Ols18}.

\section{Conclusions}
\label{con}
In this paper, we have studied the problem of generation of large-scale
magnetic field due to random flows with fixed kinetic helicity and
finite correlation times in a shearing background.
We employed a pulsed renewing flow based model where flow field renovates
itself after every time interval $\tau$, called here the correlation time
of the inviscid random helical flow. We made use of
single plane shearing waves with fixed kinetic helicity to model the
renovating flow. These are time-dependent exact solutions to the
Navier-Stokes equations as derived by \cite{SS17}. Thus we self-consistently
included the anisotropic effects of shear on the flow itself, which in turn
governs the evolution of magnetic fields. We constructed the suitable ensemble
of realization for the velocity field, later to be used in the averaging of magnetic field. By making a suitable use of shearing coordinate transformation, we wrote the ideal induction equation in a shearing frame
which translates with the background linear shear flow.
The evolution of magnetic field is determined in terms of
Cauchy's solutions in a fixed interval $\tau$ which represents a single realization. Because of discrete time-translational symmetry of renovating flow in $n\tau$, we could construct the eigenvalue problem for the mean magnetic field in Fourier space in any interval. The Green's function or the propagator (or simply average response tensor), which maps the mean-field at $(n-1)\tau$ to $n\tau$ is obtained after performing an average over
many realizations, or equivalently, over time $t=n \tau$ with $n\rightarrow \infty$.
The eigenvalues of the average response tensor determine the dispersion relation,
which yields the growth rate ($\gamma$) and cycle period ($P_{\rm cyc}$)
of growing mean-field dynamo wave. Below we first list some key properties
and assumptions of our model:

\begin{itemize}
\item Shear rate $S$ and velocity correlation time $\tau$ are the arbitrary parameters.
We have ignored the diffusion term from the induction equation in this
work to keep the analysis simpler.
\item Helical shearing waves that were used to model the renovating flow
freely evolve for the renovation time interval $\tau$, and are reseted to
the same amplitude at the beginning of each renovation interval. The parameters
of the flow take random values in different intervals such that the
velocity field becomes completely uncorrelated after time $\tau$.
Such a model of the flow tries to capture the effects of stochastic helical
forcing after every $\tau$.
\end{itemize}

We studied the properties of growth rate and cycle periods of growing
large-scale magnetic fields that are obtained by a mean-field dynamo
action due to helical stochastic flows in a background linear shear.
We focused in the regime when memory effects become important, i.e.,
when the response tensor or the turbulent EMF is affected by the
time dependence of the mean magnetic field. As clarified in \cite{SS14},
this is equivalent to the case when the random flows are correlated
for non-zero times, as in the white-noise case the generalized EMF
with a history term through a time-integral reduces to a simple expression
leading to an instantaneous relation with the mean magnetic field.
Our study thus essentially generalizes the standard
$\alpha^2 \Omega$ dynamo model to now also include the memory effects,
and tensorial nature of $\alpha$ while treating the shear
non-perturbatively.\\

\noindent
(a) \emph{Non-axisymmetric ($k_2 \neq 0$) modes:}
We found that the non-axisymmetric modes eventually decay in time and therefore are
unimportant for late time structures of mean magnetic field. However,
these modes decay in an interesting manner in that the contours of the
growth rate $\gamma$ form an ellipse with properties resembling the
one for resistive Green's function which is derived by \cite{SS10}
for nonzero $\eta$ after ignoring the advection term. As we have ignored
$\eta$ in the present work, the resemblance points to a notion of
turbulent diffusivity $\eta_t$ which typically augments $\eta$. The decay of
such non-axisymmetric modes may thus be used to determine $\eta_t$ and this
will be attempted in a future work.\\

\noindent
(b) \emph{Axisymmetric ($k_2 = 0$) modes:}
These are the only modes that survive and will determine the late time
evolution of mean magnetic field. Comparing our results of axisymmetric
mean field dynamo with the predictions of standard $\alpha \Omega$ dynamos,
we find that the behaviours of $\gamma$ and $P_{\rm cyc}$
with shear strength $|S|$ are even qualitatively different when
$\tau/T \sim {\cal O}(1)$.
Some notable findings in the regime when memory effects become important,
i.e. when $\tau/T \sim {\cal O}(1)$, are highlighted as follows:

\begin{enumerate}
\item [(i)] The growth rate $\gamma$ and the wavenumber ($k_\ast$)
corresponding to the fastest growing mode vary non-monotonically with
$|S|$; see \Figs{k3max_wS}{Relam_k3max_wSlog}.
We find the quenching of the dynamo when shear becomes sufficiently
strong. This is in agreement with the work of \cite{LK08} who also
reported dynamo quenching due to strong shear. Common notions that the
regions of strongest shear in an astrophysical object are the ideal reservoirs
of magnetic fields may thus need to be revised.

In order to understand the cause of such a quenching of growth rate ($\gamma$), we made an
attempt to determine the $\alpha$ tensor by adapting to its simplified textbook
definition. We found that the magnitude of the more relevant component
$\alpha_{22}$ is significantly suppressed at larger shear and this may have
affected the growth of mean magnetic field.

\item [(ii)] At fixed $S$ and $\tau$, $\gamma$ first increases from zero as
a function of wavenumber, reaches a maximum, and turns negative at much larger
wavenumbers. The quantity $k_\ast$ is smaller than $q$, which is the eddy
wavenumber determining the injection scale ($q^{-1}$) of kinetic energy,
for a whole range of shear. Also, at fixed shear, $k_{\ast}$ systematically
decreases when $\tau$ increases.
This promotes a genuine large-scale dynamo as magnetic fields grow maximally
at scales ($k_\ast^{-1}$) that are larger than eddy size ($q^{-1}$).

\item [(iii)] Dynamo cycle period $P_{\rm cyc}$ exhibits different scaling relations
with shear depending on the strength of the shear parameter:
$P_{\rm cyc} \propto |S|^{-1}$ when shear is small, and it becomes independent
of shear when shear becomes sufficiently strong. This is very different from
the predictions of standard $\alpha \Omega$ dynamo model which leads to a
uniform scaling, $P_{\rm cyc} \propto |S|^{-1/2}$, with shear.
\end{enumerate}

Recent observational study by \cite{Ols18} on stellar magnetic activity
cycles reveal two branches, active and inactive, in a diagnostic
activity-rotation diagram. More work is needed to fully understand
the origin of these branches, e.g., whether these trace two distinct
population of stars or are somehow related to multiple cycles from the
same star. Nevertheless, such studies emphasize the need to focus on
the dynamo cycle period $P_{\rm cyc}$ as this may have direct
implications for these observations. This motivated us to explore in
detail the properties of $P_{\rm cyc}$ in our model.
Interestingly enough, we find
two asymptotic branches when we look at the dimensionless quantity
$1/|S| P_{\rm cyc}$; see the inset of
\Fig{Pcyc_k3max_wS}. This quantity is independent of shear when shear is weak,
and varies as $|S|^{-1}$ in the strong shear regime.
It is useful to
note that the model predictions and scaling relations such as the
ones reported here are often based on kinematic analysis, whereas the
observations relate to the nonlinear stage of stellar dynamos. Therefore it is
important to investigate numerically how the scalings of $P_{\rm cyc}$ are
affected in the nonlinear stage. Nevertheless, we envisage that the
new scaling laws being reported here will be useful in the interpretations
of observations of the magnetic activity cycles of stars.

\section*{Acknowledgments}
We thank the referee for useful comments and suggestions.
We thank S. Sridhar for useful discussions and comments on the manuscript. 
We thank Axel Brandenburg and Tarun Deep Saini for their support.
NJ acknowledges the hospitality provided by Nordita, Sweden, where this work began.
NJ also thanks Chanda J. Jog for providing financial support.

\appendix
\section{Comparison with non-shearing waves of the renovating flows}
\label{unshear_comp}

We choose the turbulent velocity $\bfu$ same as in GB92, 
\beq
\bfu(\bfX,t) = \bfa\sin(\bfq\cendot\bfX + \Psi) + h\,\bfc\cos(\bfq\cendot\bfX + \Psi)
\eeq
which is a single helical waves with constant wave vector $\bfq$ and constant amplitudes $\bfa$ and $\bfc$. We call them \textsl{non-shearing waves}, since both the amplitudes and the wave vector are independent of shear. This velocity field is used instead of \Eq{vel} to obtain the growth rate ($\gamma$) and maximum growing wavenumber ($k_\ast$). Such velocity field give, isotropic transport co-efficient--$\alpha$, if we use \Eq{alpha_def}. This velocity fields are used, so that we can make the comparison between the cases, when $\alpha$ is tensorial and a function of shear, with isotropic $\alpha$--independent of shear--in this appendix. The effect of shear on the turbulent eddy is not considered in the non-shearing waves as it is usually the case in $\alpha\Omega$ $(\alpha^2\Omega)$ dynamo. 

\begin{figure}
\includegraphics[width=0.5\columnwidth]{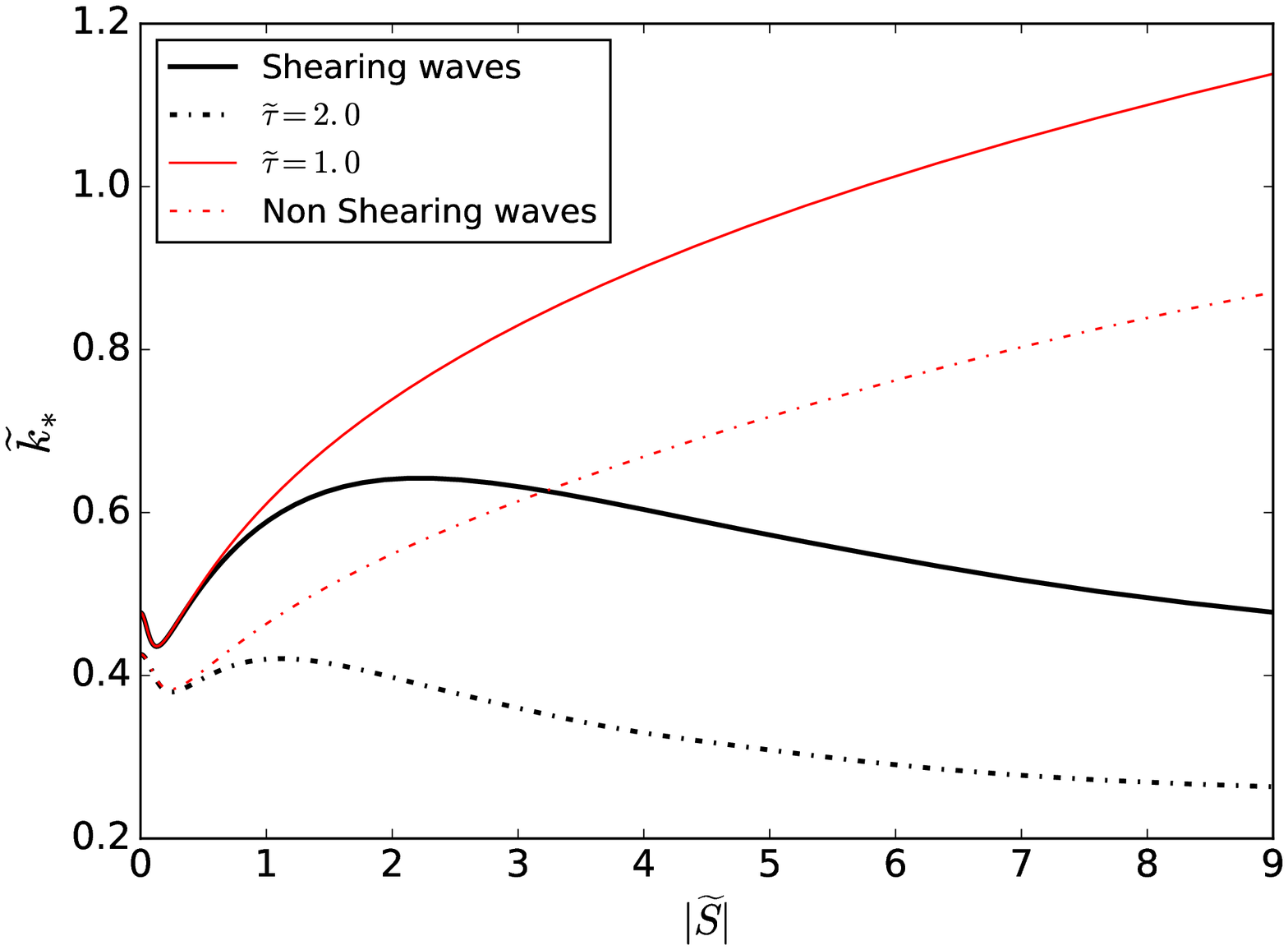}
\includegraphics[width=0.5\columnwidth]{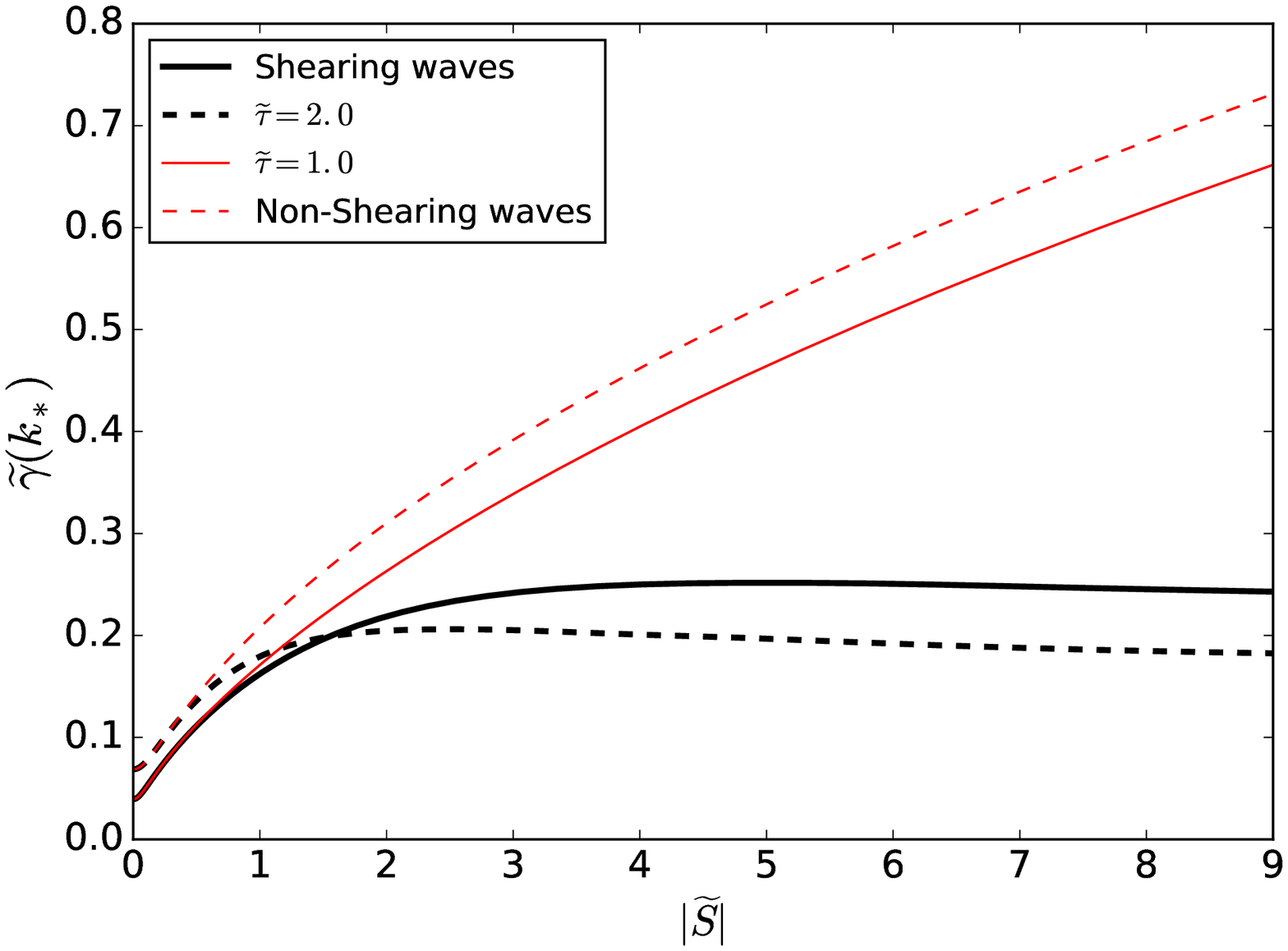}
\caption{Normalized growth rate $\tilde{\gamma}(k_\ast)$
and normalized maximum growing mode $k_\ast$ of growing dynamo
are shown in left and right panels, respectively.
Black: shearing waves; red: non-shearing waves;
$\tau/T=1$ (solid), and 2 (dashed)}.
\label{Relam_k3max_wSlog_compare}
\end{figure}

 We show such comparisons in
\Fig{Relam_k3max_wSlog_compare}
at $\tau/T=1$ and 2, when memory effects are important.
As may be seen from \Fig{Relam_k3max_wSlog_compare} (left panel) that the behaviour of the wavenumber $k_\ast$ corresponding
to the fastest growing dynamo mode is qualitatively
different. When we model renovating flows
in terms of non-shearing waves, $k_\ast$
increases with $|S|$, producing magnetic fields predominantly
at small, sub-eddy spatial scales at intermediate
to large values of shear (red lines).
However, as discussed before based on \Fig{k3max_wS},
when we consider amplitude modulated shearing waves
in the model, we find that $k_\ast$ remains small
at all shear, thus producing magnetic fields preferentially
at large, super-eddy spatial scales (black lines).
Note that the background shear operates at all times,
regardless of which one of the two, shearing or
non-shearing, waves we use to model the renovating flows.

In \Fig{Relam_k3max_wSlog_compare} (right panel) we show the
comparison for shear dependence of the
dynamo growth rate again at $\tau/T=1$ and 2. While the growth rate $\tilde{\gamma}(k_\ast)$ increases monotonically with shear strength $|S|$ when non-shearing waves are used in
calculations, it shows a saturation and even quenching at
large enough values of shear when we utilize time-dependent
shearing waves to model the renovating flows;
dynamo quenching in strong shear regime is more clearly
seen in \Fig{Relam_k3max_wSlog} where shear strength
is shown on a logarithmic axis. Here, we have considered the growth of mean field by the single eddy with and without the effect of shear on it's amplitudes along with the background flow. Therefore, the effect of shear on the turbulence is non-trivial which cannot be neglected and it is more pronounced at strong shear regime.

\label{lastpage}

\end{document}